\documentclass[%
 prfluids,
 superscriptaddress,
 amsmath,amssymb,
 aps,
onecolumn,
preprint,
titlepage,
longbibliography
]{revtex4-2}

\usepackage{graphicx}
\usepackage{dcolumn}
\usepackage{bm}
\usepackage{natbib}
\usepackage{color}
\usepackage{comment}

\usepackage{CJK}

\begin{document}

\begin{CJK*}{UTF8}{min}


\title{Analysis of anisotropic subgrid-scale stress\\ for coarse large-eddy simulation
}

\author{Kazuhiro Inagaki (稲垣 和寛)}
 \email{Department of Mechanical Engineering, Doshisha University,1-3 Tatara Miyakodani, Kyotanabe 610-0394, Japan. E-mail: kinagaki@mail.doshisha.ac.jp}
\affiliation{%
Research and Education Center for Natural Sciences, Keio University, 4-1-1 Hiyoshi, Kohoku-ku, Yokohama 223-8521, Japan
}


\author{Hiromichi Kobayashi (小林 宏充)}%
\affiliation{%
Department of Physics \& Research and Education Center for Natural Sciences, Hiyoshi Campus, Keio University, 4-1-1 Hiyoshi, Kohoku-ku, Yokohama 223-8521, Japan
}%

\date{\today}

\allowdisplaybreaks[1]

\begin{abstract}
This study discusses the necessity of anisotropic subgrid-scale (SGS) stress in large-eddy simulations (LESs) of turbulent shear flows using a coarse grid resolution. We decompose the SGS stress into two parts to observe the role of SGS stress in turbulent shear flows in addition to the energy transfer between grid-scale (GS or resolved scale) and SGS. One is the isotropic eddy-viscosity term, which contributes to energy transfer, and the other is the residual anisotropic term, which is separated from the energy transfer. We investigate the budget equation for GS Reynolds stress in turbulent channel flows accompanied by the SGS stress decomposition. In addition, we examine the medium and coarse filter length cases; the conventional eddy-viscosity models can fairly predict the mean velocity profile for the medium filter case and fails for the coarse filter case. The budget for GS turbulent kinetic energy shows that the anisotropic SGS stress has a negligible contribution to energy transfer. In contrast, the anisotropic stress has a large and non-dissipative contribution to the streamwise and spanwise components of GS Reynolds stress when the filter size is large. Even for the medium-size filter case, the anisotropic stress contributes positively to the budget for the spanwise GS Reynolds stress. Spectral analysis of the budget reveals that the positive contribution is prominent at a scale consistent with the spacing of streaks in the near-wall region. Therefore, we infer that anisotropic stress contributes to the generation mechanism of coherent structures. Predicting the positive contribution of the anisotropic stress to the budget is key to further improving SGS models.

\end{abstract}

\pacs{Valid PACS appear here}
\maketitle

\end{CJK*}


\section{\label{sec:level1}Introduction}

Subgrid-scale (SGS, or subfilter-scale) models play a significant role in accurately predicting the statistics of turbulent flows in large-eddy simulations (LESs). For instance, eddy-viscosity models provide a statistically accurate energy transfer rate from grid scale (GS or resolved scale) to SGS. However, an accurate energy transfer rate is only one of the required conditions for predicting the statistics of turbulent flows. In general, characteristic quantities other than the energy transfer rate are important in predicting turbulent flows; for example, the mean SGS shear stress in turbulent shear flows. Several studies pointed out that the eddy-viscosity models are not necessarily sufficient for predicting both the energy transfer rate and mean SGS stress in turbulent shear flows \cite{meneveau1994,jm2000,lm2004}. Therefore, it is important to understand the role of SGS stress in addition to energy transfer for the further development of LES.

Several studies addressed the development of non-eddy-viscosity or anisotropic SGS models and showed that these models improve the prediction of the statistics of turbulent flows \cite{marstorpetal2009,montecchiaetal2017,abe2013,ia2017,ik2020,agrawaletal2022,cda2014,cimarellietal2019}. Specifically, these anisotropic models provide a better prediction even in grid resolution cases that are coarser than in conventional eddy-viscosity models. Therefore, we infer that the difference between eddy-viscosity-based and anisotropic models becomes prominent in coarse grid cases. Analysis using a coarse grid or large filter scale will shed light on the role of anisotropic SGS stress in LES. Furthermore, the necessity of improving the SGS model in a coarse grid resolution has recently been discussed for atmospheric turbulence as the grey zone or terra incognita problem \cite{honnertetal2020}.

It is worth noting that some anisotropic models prohibit backward scatter, which is the local energy transfer from SGS to GS \cite{marstorpetal2009,montecchiaetal2017,abe2013,ia2017,ik2020,agrawaletal2022}. This indicates that additional stress, apart from energy transfer, can improve the performance of LES. Actually, Abe \cite{abe2019} demonstrated that the additional stress apart from the energy transfer essentially contributes to the generation of GS or resolved scale Reynolds shear stress in turbulent channel flows. Based on a similar analysis, Inagaki and Kobayashi \cite{ik2020} suggested that the amplification of small-scale velocity fluctuations close to the cut-off scale due to the anisotropic stress contributes to the improvement of the prediction of statistics in LES. As suggested by these studies, an analysis of the budget for the Reynolds stress will shed light on the physical role of anisotropic SGS stress in a statistical sense.

In this study, we investigate the budget for GS Reynolds stress in turbulent channel flows. To clarify the physical role of the anisotropy of SGS stress, we decompose the SGS stress according to the description by Abe \cite{abe2019}. The eddy-viscosity coefficient is determined by assuming that all energy transfers between GS and SGS are governed by the eddy-viscosity term. Residual stress is obtained by subtracting the eddy-viscosity term from the exact SGS stress. Therefore, the residual stress is apart from the energy transfer. Because the eddy-viscosity approximation is often referred to as the isotropic model owing to the scalar coefficient, the residual stress represents the anisotropy of the turbulent velocity fluctuation in SGS. Such anisotropic stress is essential for predicting the anisotropy of the SGS dissipation tensor (hereafter often referred to as SGS dissipation, simply), which is the key ingredient in this study.

The anisotropy of SGS dissipation is rarely considered in the context of SGS modeling \cite{moseretal2021}. Haering \textit{et al.} \cite{haeringetal2019} demonstrated that a simple tensor-coefficient eddy-viscosity model can improve the prediction of SGS dissipation anisotropy induced by anisotropic grids. Marstorp \textit{et al.} \cite{marstorpetal2009} reported that LES employing anisotropic SGS stress improves the prediction of the ratio of wall-normal to streamwise components of SGS dissipation when compared with the eddy-viscosity model in rotating channel flows. In conventional non-rotating channel flow, Domaradzki \textit{et al.} \cite{domaradzkietal1994} and H{\"a}rtel and Kleiser \cite{hk1998} showed that in the energy transfer between GS and SGS, the streamwise component is dominant and others are negligible in the near-wall region. As shown later, the anisotropic property of SGS dissipation cannot be reproduced solely by the eddy-viscosity term even though the total SGS dissipation, which is the trace part of SGS dissipation tensor, is described adequately by the isotropic eddy viscosity. Therefore, this study quantifies the importance of the anisotropic SGS stress in predicting the statistics of turbulent flows in LES.

The rest of this paper is organized as follows. First, we summarize the decomposition of SGS stress according to the description by Abe \cite{abe2019} and budget equation for GS Reynolds stress in Sec.~\ref{sec:level2}. We also provide a physical interpretation of SGS stress-related terms in the budget. The numerical results for the budgets in turbulent channel flows are presented in Sec.~\ref{sec:level3}. In Sec.~\ref{sec:level4}, further analysis of the budgets in Fourier space is presented to discuss the physical role of anisotropic SGS stress in the near-wall region. In addition, we provide an \textit{a priori} test of the SGS dissipation based on anisotropic SGS stress using a few existing model expressions. Conclusions are presented in Sec~\ref{sec:level5}.

\section{\label{sec:level2}Budget equation for GS Reynolds stress}

\subsection{\label{sec:level2a}Decomposition of SGS stress}

In LES of incompressible flows, the governing equations are filtered continuity and Navier-Stokes equations:
\begin{gather}
\frac{\partial \overline{u}_i}{\partial x_i} = 0, 
\label{eq:1} \\
\frac{\partial \overline{u}_i}{\partial t} = - \frac{\partial}{\partial x_j} ( \overline{u}_i \overline{u}_j + \tau^\mathrm{sgs}_{ij} ) - \frac{\partial \overline{p}}{\partial x_i} + \frac{\partial}{\partial x_j} (2\nu \overline{s}_{ij}),
\label{eq:2} 
\end{gather}
where $\overline{\cdot}$ represents the spatial filtering 
operation. Here, we assume that the filter and partial differential operations are always commutative. If the filter and partial differential operations are not commutative, several additional terms appear in the filtered continuity and Navier--Stokes equations (see e.g. Ref.~\cite{moseretal2021}), which makes the investigation of the effects of SGS on GS much more complex. 
To focus on the role of SGS stress in the governing equations, we assume the commutativity of filter operation. 
$\overline{u}_i$, $\overline{p}$, and $\overline{s}_{ij} [ = ( \partial \overline{u}_i/\partial x_j + \partial \overline{u}_j/\partial x_i)/2]$ are the GS velocity, pressure divided by the density, and strain-rate tensor, respectively. $\nu$ is the kinematic viscosity. The sole unknown variable in LES is the SGS stress defined by $\tau^\mathrm{sgs}_{ij} = \overline{u_i u_j} - \overline{u}_i \overline{u}_j$. 

To numerically solve Eqs.~(\ref{eq:1}) and (\ref{eq:2}), we must employ a closed model expression for $\tau^\mathrm{sgs}_{ij}$. In contrast, in the analysis based on direct numerical simulation (DNS) data, which is referred to as the \textit{a priori} test, we can directly calculate $\tau^\mathrm{sgs}_{ij}$ by explicit filtering. However, in the case of an \textit{a priori} test, we cannot assess the amount of SGS stress that can be modeled in terms of the eddy-viscosity approximation. The inconsistency between the \textit{a priori} and \textit{a posteriori} tests of the SGS model increases the complexity of this issue. The correlation between SGS stress and strain rate in the \textit{a priori} test is low \cite{liuetal1994,taoetal2002,horiuti2003}. Furthermore, abundant backward scatter events \cite{piomellietal1991,aoyamaetal2005} reject the validity of an eddy-viscosity model with a positive coefficient. Nevertheless, the purely dissipative eddy-viscosity models perform well in several turbulent flows in the \textit{a posteriori} test of LES (see Ref. \cite{moseretal2021}).

The eddy-viscosity term is ideal to achieve numerical stability, although its justification is still under discussion. In particular, when we employ the scale-similarity model, the eddy-viscosity term is often employed to compensate for insufficient dissipation \cite{bardinaetal1983,mk2000}. Furthermore, recent studies on anisotropic SGS modeling demonstrated that the employment of the additional stress that has no contribution to the energy transfer improves the prediction of the statistics in LES \cite{marstorpetal2009,montecchiaetal2017,abe2013,ia2017,ik2020,agrawaletal2022}. Therefore, we adopt the following decomposition of SGS stress by assuming that the eddy-viscosity term governs the energy transfer between GS and SGS \cite{abe2019,ik2020}:
\begin{align}
\tau^\mathrm{sgs}_{ij} - \frac{1}{3} \tau^\mathrm{sgs}_{\ell \ell} \delta_{ij} = 
- 2 \nu^\mathrm{sgs} \overline{s}_{ij} + \tau^\mathrm{ani}_{ij},
\label{eq:3}
\end{align}
where
\begin{gather}
\nu^\mathrm{sgs} = - \frac{\tau^\mathrm{sgs}_{ij} \overline{s}_{ij}}{2 \overline{s}^2},
\label{eq:4} \\
\tau^\mathrm{ani}_{ij} = \tau^\mathrm{sgs}_{ij} - \frac{1}{3} \tau^\mathrm{sgs}_{\ell \ell} \delta_{ij}
+ 2 \nu^\mathrm{sgs} \overline{s}_{ij},
\label{eq:5}
\end{gather}
and $\overline{s}^2 = \overline{s}_{ij} \overline{s}_{ij}$. According to Abe \cite{abe2019}, we refer to $\tau^\mathrm{ani}_{ij}$ as the extra anisotropic or simply the anisotropic term. Based on the definition of eddy viscosity given by Eq.~(\ref{eq:4}), the anisotropic term does not contribute to the energy transfer between GS and SGS, that is, $\tau^\mathrm{ani}_{ij} \overline{s}_{ij} = 0$. Although this decomposition is arbitrary, this analysis allows us to verify the physical properties of the anisotropic part of SGS stress, in addition to the energy transfer. The eddy-viscosity defined by Eq.~(\ref{eq:4}) can be negative, which indicates the local backward scatter of kinetic energy. However, the average energy transfer rate is almost positive because of the predominance of forward scatter in turbulent flows. It is worth noting that this eddy-viscosity does not necessarily predict the mean SGS shear stress in turbulent shear flows even though it accounts for an accurate energy transfer rate \cite{meneveau1994,jm2000,lm2004}. Hence, the present analysis also provides the difference between the exact mean SGS shear stress and that predicted only by the eddy-viscosity term based on energy transfer.

\subsection{\label{sec:level2b}Budget equation for GS Reynolds stress}

Under Reynolds decomposition, $\langle \overline{q} \rangle = Q$ and $\overline{q}' = \overline{q} - \langle \overline{q} \rangle = \overline{q} - Q$ for $q = (u_i,p,s_{ij})$ with the ensemble average $\langle \cdot \rangle$,
the budget equation for GS Reynolds stress $R^\mathrm{GS}_{ij} (=\langle \overline{u}_i' \overline{u}_j' \rangle)$ in the LES reads
\begin{align}
\frac{\partial R^\mathrm{GS}_{ij}}{\partial t} + \frac{\partial}{\partial x_\ell} (U_\ell R^\mathrm{GS}_{ij})
= P^\mathrm{GS}_{ij} - \varepsilon^\mathrm{GS}_{ij} + \Phi^\mathrm{GS}_{ij}
+ D^\mathrm{p,GS}_{ij} + D^\mathrm{t,GS}_{ij} + D^\mathrm{v,GS}_{ij}
- \varepsilon^\mathrm{SGS}_{ij} + D^\mathrm{SGS}_{ij},
\label{eq:6}
\end{align}
where
\begin{subequations}
\begin{align}
P^\mathrm{GS}_{ij} & =
- R^\mathrm{GS}_{i\ell} \frac{\partial U_j}{\partial x_\ell} - R^\mathrm{GS}_{j\ell} \frac{\partial U_i}{\partial x_\ell}, 
\label{eq:7a} \\
\varepsilon^\mathrm{GS}_{ij} & = 2 \nu \left< \overline{s}_{i\ell}' \frac{\partial \overline{u}_j'}{\partial x_\ell} + \overline{s}_{j\ell}' \frac{\partial \overline{u}_i'}{\partial x_\ell} \right>, 
\label{eq:7b} \\ 
\Phi^\mathrm{GS}_{ij} & = 2 \left< \overline{p}^\mathrm{total} \overline{s}_{ij}' \right>, 
\label{eq:7c} \\ 
D^\mathrm{p,GS}_{ij} & = - \frac{\partial}{\partial x_\ell} \left< \overline{p}^\mathrm{total} \overline{u}_j' \delta_{i\ell} + \overline{p}^\mathrm{total} \overline{u}_j' \delta_{i\ell} \right>, 
\label{eq:7d} \\ 
D^\mathrm{t,GS}_{ij} & = -\frac{\partial}{\partial x_\ell} \left< \overline{u}_\ell' \overline{u}_i' \overline{u}_j' \right>, 
\label{eq:7e} \\ 
D^\mathrm{v,GS}_{ij} & = 2\nu \frac{\partial}{\partial x_\ell} \left< \overline{s}_{i\ell}' \overline{u}_j' + \overline{s}_{j\ell}' \overline{u}_j' \right>,
\label{eq:7f} \\
\varepsilon^\mathrm{SGS}_{ij} & = -\left< \tau^\mathrm{sgs}_{i\ell}|_\mathrm{tl} \frac{\partial \overline{u}_j'}{\partial x_\ell}
+ \tau^\mathrm{sgs}_{j\ell}|_\mathrm{tl} \frac{\partial \overline{u}_i'}{\partial x_\ell} \right>,
\label{eq:7g} \\ 
D^\mathrm{SGS}_{ij} & = - \frac{\partial}{\partial x_\ell} \left< \tau^\mathrm{sgs}_{i\ell}|_\mathrm{tl} \overline{u}_j' + \tau^\mathrm{sgs}_{j\ell}|_\mathrm{tl} \overline{u}_i' \right>,
\label{eq:7h}
\end{align}
\end{subequations}
$\overline{p}^\mathrm{total} = \overline{p} + \tau^\mathrm{sgs}_{mm} /3$, and $\tau^\mathrm{sgs}_{ij}|_\mathrm{tl} = \tau^\mathrm{sgs}_{ij} - \tau^\mathrm{sgs}_{mm} \delta_{ij}/3$. The terms expressed in Eqs.~(\ref{eq:7a})-(\ref{eq:7h}) are referred to as production, dissipation, pressure redistribution, pressure diffusion, turbulent diffusion, viscous diffusion, SGS dissipation, and SGS diffusion, respectively. Contributions from SGS stress appear in the SGS dissipation $\varepsilon^\mathrm{SGS}_{ij}$ and diffusion $D^\mathrm{SGS}_{ij}$. Note that SGS dissipation is not necessarily positive, even for the trace part $\varepsilon^\mathrm{SGS} (= \varepsilon^\mathrm{SGS}_{ii}/2)$, in contrast to the molecular dissipation $\varepsilon (= \varepsilon_{ii}/2)$. Negative SGS dissipation is observed near the wall in turbulent channel flows \cite{domaradzkietal1994,harteletal1994,hk1998,cda2012}. In this study, we do not focus on modeling the eddy viscosity, which predicts the spatial profile of SGS dissipation. Instead, we extract the physical properties required for anisotropic SGS stress by the decomposition (\ref{eq:3}) with (\ref{eq:4}). This analysis demonstrates the general property of anisotropic SGS stress; that is, it does not rely on a specific model expression.

\subsection{\label{sec:level2c}Decomposition of SGS dissipation}

Using Eq.~(\ref{eq:3}), we decompose the SGS dissipation $\varepsilon^\mathrm{SGS}_{ij}$ into the following forms:
\begin{align}
\varepsilon^\mathrm{SGS}_{ij} & = \varepsilon^\mathrm{EV}_{ij} - \xi^\mathrm{AR}_{ij}, 
\label{eq:8}
\end{align}
where
\begin{subequations}
\begin{align}
\varepsilon^\mathrm{EV}_{ij} & = 2 \left< \nu^\mathrm{sgs} \left( \overline{s}_{i\ell} \frac{\partial \overline{u}_j'}{\partial x_\ell}
+ \overline{s}_{j\ell} \frac{\partial \overline{u}_i'}{\partial x_\ell} \right) \right>,
\label{eq:9a} \\
\xi^\mathrm{AR}_{ij} & = \left< \tau^\mathrm{ani}_{i\ell} \frac{\partial \overline{u}_j'}{\partial x_\ell}
+ \tau^\mathrm{ani}_{j\ell} \frac{\partial \overline{u}_i'}{\partial x_\ell} \right>.
\label{eq:9b} 
\end{align}
\end{subequations}
We refer to the terms defined in Eqs.~(\ref{eq:9a}) and (\ref{eq:9b}) as eddy-viscosity dissipation and anisotropic redistribution, respectively. The eddy-viscosity dissipation $\varepsilon^\mathrm{EV}_{ij}$ (\ref{eq:9a}) has a form similar to that of the conventional molecular dissipation $\varepsilon_{ij}$ (\ref{eq:7b}). Hence, we expect that the eddy-viscosity dissipation plays a dissipative role similar to the molecular one. In contrast, the anisotropic redistribution $\xi^\mathrm{AR}_{ij}$ (\ref{eq:9b}) plays a different role in contrast to the eddy-viscosity dissipation (\ref{eq:9a}) owing to the anisotropy of SGS stress. Therefore, the anisotropy of SGS stress plays a significant role in predicting turbulent flows when the anisotropic redistribution contributes significantly to the GS Reynolds stress budget. Similarly, we can decompose the SGS diffusion $D^\mathrm{SGS}_{ij}$ (\ref{eq:7h}) into eddy-viscosity and anisotropic parts. For simplicity, we discuss the result of the decomposition of SGS dissipation. The decomposition of SGS diffusion is briefly discussed in Appendix~\ref{sec:a}.

The naming of anisotropic redistribution is based on the property that $\tau^\mathrm{ani}_{ij}$ has no contribution to the energy transfer between GS and SGS. In addition, we do not refer to this term as anisotropic dissipation because we expect it to be more prominently non-dissipative than SGS dissipation $\varepsilon^\mathrm{SGS}_{ij}$. Exactly speaking, however, the anisotropic redistribution does not redistribute the energy among the normal stress components, in contrast to the pressure redistribution $\Phi^\mathrm{GS}_{ij}$ (\ref{eq:7c}). In other words, the trace of pressure redistribution is exactly zero $\Phi^\mathrm{GS}_{ii} = 2 \langle \overline{p}^\mathrm{total} \overline{s}_{ii}' \rangle = 0$, whereas the anisotropic redistribution is not necessarily traceless; that is, $\xi^\mathrm{AR}_{ii} = 2 \langle \tau^\mathrm{ani}_{ij} \overline{s}_{ij}' \rangle \neq 0$ \cite{ik2020}. This non-zero trace emanates from the Reynolds decomposition; that is,
\begin{align}
\left< \tau^\mathrm{ani}_{ij} \overline{s}_{ij} \right> = \left< \tau^\mathrm{ani}_{ij} \right> S_{ij} + \xi^\mathrm{AR}_{ii}/2 = 0,
\label{eq:10}
\end{align}
and hence $\xi^\mathrm{AR}_{ii} = - \langle \tau^\mathrm{ani}_{ij} \rangle S_{ij} \neq 0$. Here, we used the property $\tau^\mathrm{ani}_{ij} \overline{s}_{ij}=0$, which is provided in Sec.~\ref{sec:level2a}. $\langle \tau^\mathrm{ani}_{ij} \rangle S_{ij}$ represents the energy transfer rate between the mean and SGS fields due to the anisotropic stress. For unidirectional turbulent shear flows as channel flows, it reads $\langle \tau^\mathrm{ani}_{xy} \rangle \partial U_x/\partial y/2$. When $\partial U_x/\partial y > 0$ and $\langle \tau^\mathrm{ani}_{xy} \rangle < 0$ as the conventional turbulent shear stress in shear flows, $\xi^\mathrm{AR}_{ii} > 0$, thus implying that the trace of anisotropic redistribution is productive. Conversely, when $\partial U_x/\partial y > 0$ and $\langle \tau^\mathrm{ani}_{xy} \rangle > 0$, $\xi^\mathrm{AR}_{ii} < 0$, thus implying that the trace of anisotropic redistribution is dissipative.

\subsection{\label{sec:level2d}Budget equation for GS Reynolds stress spectrum}

For a further understanding of SGS modeling, an analysis of the energy budget in scale space is useful. Domaradzki \textit{et al.} \cite{domaradzkietal1994} showed that in turbulent channel flows, the local energy transfer in scale space through the nonlocal wavenumber triad interaction is the majority in the interscale energy transfer process. This result suggests that the interscale interaction across the cutoff scale plays a significant role in energy transfer, which should be implemented in SGS stress. As another approach to investigating interscale interaction, Cimarelli and De Angelis \cite{cda2012} analyzed the Kolmogorov equation, which is the budget equation for the second-order velocity structure function, for both unfiltered and filtered velocity fields. They concluded that to resolve the interscale interaction including the inverse cascade observed in the spanwise scale, the appropriate filter length scale for eddy-viscosity models yields 
$\overline{\Delta}_x^+ < 100$ and $\overline{\Delta}_z^+ < 20$ where $\overline{\Delta}_i$ denotes the filter length scale for the $i$-th direction, $x$ and $z$ are the streamwise and spanwise directions, and the values with a superscript $+$ denote those normalized by the wall shear stress and kinematic viscosity.
In other words, we have to employ additional stress to the eddy-viscosity term when using a larger filter length in LES. This limitation of the filter length scale is consistent with the typical grid resolution for conventional LES employing eddy-viscosity models, 
$\overline{\Delta}_x^+ < 130$ and $\overline{\Delta}_z^+ < 30$ (see Refs.~\cite{kravchenkoetal1996,mv2001,cm2012}). 

To investigate the physical role of anisotropic stress and the limitations of eddy-viscosity models, we also analyze the budget equation for GS Reynolds stress spectrum. We consider the case in which the $x$ and $z$ directions are periodic and the $y$ direction is bounded by solid walls as turbulent channel flows. We adopt the Fourier transformation in the $x$ and $z$ directions for the scale decomposition of the flow fields; the discrete Fourier transformation of a quantity $q$ and its inverse transformation read
\begin{subequations}
\begin{align}
\tilde{q} (n_x, y, n_z) & = \frac{1}{N_x N_z} \sum_{(I,K)=(1,1)}^{(N_x, N_z)}  q (x_I, y,z_K) \exp [ - 2 \pi \mathrm{i} (n_x x_I/L_x  + n_z z_K/L_z)],
\label{eq:11a} \\
q (x_I, y, z_K) & = \sum_{(n_x,n_z) = (- N_x/2 , -N_z/2)}^{ (N_x/2-1, N_z/2-1)} \tilde{q} (n_x, y,n_z) \exp [ 2 \pi \mathrm{i} (n_x x_I/L_x  + n_z z_K/L_z)],
\label{eq:11b}
\end{align}
\end{subequations}
where $\mathrm{i} = \sqrt{-1}$, $x_I =L_x I/N_x$, and $z_K = L_z K/N_z$. $L_i$ and $N_i$ are the length of the flow domain and grid number in the $i$-th direction, respectively. When the turbulence field is statistically homogeneous in the $x$ and $z$ directions, the second-order correlation yields $\langle f(x_I,y,z_K) g (x_I,y,z_K) \rangle = \langle f g \rangle (y) = \sum_{n_x,n_z} \Re \langle \tilde{f} (n_x,y,n_z) \tilde{g}^* (n_x,y,n_z) \rangle$ where the superscript $*$ represents the complex conjugate. Hence, the GS Reynolds stress spectrum $E^\mathrm{GS}_{ij} (k_x,y,k_z)$ reads
\begin{align}
E^\mathrm{GS}_{ij} (k_x,y,k_z) = \Re \left< \tilde{\overline{u}}_i' \tilde{\overline{u}}_i'{}^* \right>,
\label{eq:12}
\end{align}
which satisfies $R^\mathrm{GS}_{ij} (y) = \sum_{n_x,n_z} E^\mathrm{GS}_{ij} (k_x,y,k_z)$ where $k_x = 2\pi n_x/L_x$ and $k_z = 2\pi n_z/L_z$.

The extension of the spectral budget for Reynolds stress \cite{ka2018,lm2019} to the filtered Navier-Stokes equations yields \cite{ik2020}
\begin{align}
\frac{\partial E^\mathrm{GS}_{ij}}{\partial t} 
= \check{P}^\mathrm{GS}_{ij} - \check{\varepsilon}^\mathrm{GS}_{ij} + \check{\Phi}^\mathrm{GS}_{ij}
+ \check{D}^\mathrm{p,GS}_{ij} + \check{D}^\mathrm{t,GS}_{ij} + \check{D}^\mathrm{v,GS}_{ij}
- \check{\varepsilon}^\mathrm{SGS}_{ij} + \check{D}^\mathrm{SGS}_{ij}
+ \check{T}^\mathrm{GS}_{ij},
\label{eq:13}
\end{align}
where
\begin{subequations}
\begin{align}
\check{P}^\mathrm{GS}_{ij} & =
- E^\mathrm{GS}_{iy} \frac{\partial U_j}{\partial y} - E^\mathrm{GS}_{jy} \frac{\partial U_i}{\partial y}, 
\label{eq:14a} \\
\check{\varepsilon}^\mathrm{GS}_{ij} & = 2 \nu \Re \left< \tilde{\overline{s}}_{i\ell}' (\tilde{\partial}_\ell \tilde{\overline{u}}_j')^* + \tilde{\overline{s}}_{j\ell}' (\tilde{\partial}_\ell \tilde{\overline{u}}_i')^* \right>, 
\label{eq:14b} \\ 
\check{\Phi}^\mathrm{GS}_{ij} & = 2 \Re \left< \tilde{\overline{p}}^\mathrm{total} \tilde{\overline{s}}_{ij}'{}^* \right>, 
\label{eq:14c} \\ 
\check{D}^\mathrm{p,GS}_{ij} & = - \frac{\partial}{\partial y} \Re \left< \tilde{\overline{p}}^\mathrm{total} \tilde{\overline{u}}_j'{}^* \delta_{iy} + \tilde{\overline{p}}^\mathrm{total} \tilde{\overline{u}}_j'{}^* \delta_{iy} \right>, 
\label{eq:14d} \\ 
\check{D}^\mathrm{t,GS}_{ij} & = - \frac{1}{2} \frac{\partial}{\partial y} \Re \left< \widetilde{\overline{u}_y' \overline{u}_i'} \tilde{\overline{u}}_j'{}^* + \widetilde{\overline{u}_y' \overline{u}_j'} \tilde{\overline{u}}_i'{}^* \right>, 
\label{eq:14e} \\ 
\check{D}^\mathrm{v,GS}_{ij} & = 2\nu \frac{\partial}{\partial x_y} \Re \left< \tilde{\overline{s}}_{iy}' \tilde{\overline{u}}_j'{}^* + \tilde{\overline{s}}_{jy}' \tilde{\overline{u}}_j'{}^* \right>,
\label{eq:14f} \\
\check{\varepsilon}^\mathrm{SGS}_{ij} & = - \Re \left< \tilde{\tau}^\mathrm{sgs}_{i\ell}|_\mathrm{tl} (\tilde{\partial}_\ell \tilde{\overline{u}}_j')^*
+ \tilde{\tau}^\mathrm{sgs}_{j\ell}|_\mathrm{tl} (\tilde{\partial}_\ell \tilde{\overline{u}}_i')^* \right>,
\label{eq:14g} \\ 
\check{D}^\mathrm{SGS}_{ij} & = - \frac{\partial}{\partial y} \Re \left< \tilde{\tau}^\mathrm{sgs}_{iy}|_\mathrm{tl} \tilde{\overline{u}}_j'{}^* + \tilde{\tau}^\mathrm{sgs}_{j\ell}|_\mathrm{tl} \tilde{\overline{u}}_i'{}^* \right>,
\label{eq:14h} \\
\check{T}^\mathrm{GS}_{ij} & = \Re \left< \tilde{N}_i \tilde{\overline{u}}_j'{}^* + \tilde{N}_j \tilde{\overline{u}}_i'{}^* \right> - \check{P}^\mathrm{GS}_{ij} - \check{D}^\mathrm{t,GS}_{ij},
\label{eq:14i}
\end{align}
\end{subequations}
$(\tilde{\partial}_x,\tilde{\partial}_y,\tilde{\partial}_z) = (\mathrm{i} k_x, \partial/\partial y,\mathrm{i} k_z)$, and $N_i = -\partial \overline{u}_i \overline{u}_\ell/\partial x_\ell$. Here, we assumed homogeneity of the turbulence field in the $x$ and $z$ directions. All terms on the right-hand side of Eq~(\ref{eq:13}) except for $\check{T}^\mathrm{GS}_{ij}$ lead to the right-hand side of Eq.~(\ref{eq:6}) when summed over the wavenumber; that is,
\begin{align}
A_{ij} (y) = \sum_{(n_x,n_z) = (-N_x/2,-N_z/2)}^{(N_x/2-1,N_z/2-1)} \check{A}_{ij} (k_x,y,k_z),
\label{eq:15}
\end{align}
for $A_{ij} = (P^\mathrm{GS}_{ij},\varepsilon^\mathrm{GS}_{ij},\Phi^\mathrm{GS}_{ij},D^\mathrm{p,GS}_{ij},D^\mathrm{t,GS}_{ij},D^\mathrm{v,GS}_{ij},\varepsilon^\mathrm{SGS}_{ij},D^\mathrm{SGS}_{ij})$. We refer to these terms with the same names as those given in Eqs.~(\ref{eq:7a})--(\ref{eq:7h}). However, the sum of $\check{T}^\mathrm{GS}_{ij}$ over the wavenumber yields zero; that is,
\begin{align}
\sum_{(n_x,n_z) = (-N_x/2,-N_z/2)}^{(N_x/2-1,N_z/2-1)} \check{T}_{ij} (k_x,y,k_z) = 0.
\label{eq:16}
\end{align}
Therefore, $\check{T}^\mathrm{GS}_{ij}$ represents the transfer of $E^\mathrm{GS}_{ij}$ among the wavenumber modes. We refer to $\check{T}^\mathrm{GS}_{ij}$ as the interscale transfer term.

For the spectral expressions of SGS dissipation, we also employ the decomposition provided in Sec.~\ref{sec:level2c}. Namely, 
\begin{align}
\check{\varepsilon}^\mathrm{SGS}_{ij} & = \check{\varepsilon}^\mathrm{EV}_{ij} - \check{\xi}^\mathrm{AR}_{ij}, 
\label{eq:17}
\end{align}
where
\begin{subequations}
\begin{align}
\check{\varepsilon}^\mathrm{EV}_{ij} & = 2 \left< \widetilde{\nu^\mathrm{sgs} \overline{s}_{i\ell}} (\tilde{\partial}_\ell \tilde{\overline{u}}_j')^*
+ \widetilde{\nu^\mathrm{sgs} \overline{s}_{j\ell}} (\tilde{\partial}_\ell \tilde{\overline{u}}_i')^* \right>,
\label{eq:18a} \\
\check{\xi}^\mathrm{AR}_{ij} & = \left< \tilde{\tau}^\mathrm{ani}_{i\ell} (\tilde{\partial}_\ell \tilde{\overline{u}}_j')^*
+ \tilde{\tau}^\mathrm{ani}_{j\ell} (\tilde{\partial}_\ell \tilde{\overline{u}}_i')^* \right>.
\label{eq:18b}
\end{align}
\end{subequations}
Both terms on the right-hand side of Eqs.~(\ref{eq:17}) obey Eq.~(\ref{eq:15}) and are referred to as eddy-viscosity dissipation and anisotropic redistribution, respectively.

\section{\label{sec:level3}Results in turbulent channel flows}

\subsection{\label{sec:level3a}Numerical setup}

We investigate the budget equation for GS Reynolds stress provided in Sec.~\ref{sec:level2b} using the DNS data of the turbulent channel flows. The $x$, $y$, and $z$ coordinates represent the streamwise, wall-normal, and spanwise directions, respectively. We employed a staggered grid system and adopted a fourth-order conservative central finite difference scheme for the $x$ and $z$ directions and a second-order conservative finite difference scheme on the non-uniform grid for the $y$ direction \cite{kajishimatairabook}. Periodic boundary conditions were employed in the $x$ and $z$ directions and no-slip condition was employed in the $y$ direction. The second-order Adams-Bashforth scheme was employed, for time marching. The Poisson equation for pressure was solved using a fast Fourier transformation. The Reynolds number is set to be $\mathrm{Re}_\tau = 400$ where $\mathrm{Re}_\tau = u_\tau h/\nu$, $u_\tau (= \sqrt{\nu |\partial U_x/\partial y|_\mathrm{wall} |})$ is the friction velocity, and $h$ is the channel half width. The computational domain size is $L_x \times L_y \times L_z = 2\pi h \times 2h \times \pi h$ and the number of grid is $N_x \times N_y \times N_z = 256 \times 192 \times 256$. The numerical resolutions in each direction are $\Delta x^+ = 9.8$, $\Delta y^+ = 0.34$-$10$, and $\Delta z^+ = 4.9$, respectively. Here and hereafter, the values with a superscript $+$ denote those normalized by $u_\tau$ and $\nu$. The statistical average is obtained over the $x$-$z$ plane and time. The validation of our simulation is discussed in Appendix~\ref{sec:b}.

To calculate the filtered quantities, we adopted a Fourier sharp-cut filter, which is commutative with a differential operation. Note that in the calculus of GS and SGS dissipation spectra, Eqs.~(\ref{eq:14b}) and (\ref{eq:14g}), we have to use the modified wavenumber for $\tilde{\partial}_i$ owing to employing a finite difference scheme \cite{kajishimatairabook}; namely, for the fourth-order central finite difference, $k_x^\mathrm{mod} = [27 \sin (\pi n_x/N_x)-\sin (3\pi n_x/N_x)]/(12\Delta x)$. For simplicity, the filter operation is applied only in the $x$ and $z$ directions. 
In the actual LES or \textit{a posteriori} tests, the filter in the wall-normal direction is often applied because the grid size along the wall-normal direction is coarser than that of DNS. Furthermore, the scale of eddies in the wall-normal direction is also important to discuss the dynamics of wall-bounded turbulent flows \cite{cimarellietal2016}. Thus, the filter in the wall-normal direction can provide physical insight into wall-bounded turbulent flows. 
However, if we realize the filter that coincides with the actual LES, we have to apply the inhomogeneous filter where the filter length changes against the distance from the wall. This inhomogeneous filter is not commutative with a differential operation and induces several additional terms which are different from the stress term in the filtered continuity and Navier-Stokes equations \cite{moseretal2021}. In the conventional LES, we often ignore such terms arising from the commutation error and model only the SGS stress or related numerical viscosity. In this study, we focus on the role of SGS stress in the budget for GS Reynolds stress and its spectrum, which should be reproduced in the actual LES. For this reason, we apply the filter only in the $x$ and $z$ directions.
When the filter is applied only in homogeneous directions, the statistical average of the filtered variable yields the same as that of the unfiltered variable, namely, $\langle \overline{q} \rangle = \langle q \rangle$. Owing to this property, the second moment can be straightforwardly decomposed into three of mean, GS turbulence, and SGS parts. For example, the statistical average of kinetic energy $\langle u_i u_i \rangle/2$ is decomposed into mean $U_i U_i/2$, GS turbulence $K^\mathrm{GS} (=\langle \overline{u}_i' \overline{u}_i' \rangle/2)$, and SGS $\langle \overline{u_i u_i} - \overline{u}_i \overline{u}_i \rangle/2$. 
If we also apply the filtering in the wall-normal direction, we have $\langle \overline{q} \rangle \neq \langle q \rangle$, and thus the mean of filtered velocity $\langle \overline{u}_x \rangle$ is somewhat smoothed compared with the mean velocity of DNS $\langle u_x \rangle$. However, in the near-wall to buffer region, LES often predicts the mean velocity profile well. Therefore, we infer that the interaction between the mean flow and GS turbulent field through the mean velocity gradient can be reasonably discussed to some extent even if we apply the filter only in the $x$ and $z$ directions.
To observe the filter-size dependence, we employed two filter sets, namely, medium-size filter where the set of cutoff wavelength scales is $(\lambda^{\mathrm{c}+}_x, \lambda^{\mathrm{c}+}_z) = (105, 52.4)$ and coarse- or large-size filter where $(\lambda^{\mathrm{c}+}_x, \lambda^{\mathrm{c}+}_z) = (209, 105)$. Here $\lambda_\alpha = 2\pi/k_\alpha$ for $\alpha = x,z$ that correspond to LES resolutions of $(\Delta x^+, \Delta z^+) = (52.4, 26.2)$ and $(105, 52.4)$, respectively. 
Strictly speaking, an exact cut-off wavelength is often slightly shorter than the length of grid size because of dealiasing in a spectral scheme or the accuracy of the discretization scheme. For example for a fourth-order central finite difference scheme, spatial differentiation yields the modified wavenumber mentioned above whose modulus is smaller than the exact wavenumber. In this study, we do not consider such an effective maximum wavenumber to simplify the interpretation of the analysis.
According to the typical resolutions referred to in Sec.~\ref{sec:level2d}, the eddy-viscosity models can predict the basic statistics for the medium filter case, whereas the large filter case is outside their range (see also Appendix~\ref{sec:c}). Hence, by comparing the results of these two filter sizes, we can determine the physical role of the anisotropic SGS stress.

\subsection{\label{sec:level3b}Filter wavelength and GS Reynolds stresses}

\begin{figure}[tb]
\centering
\centering
\includegraphics[width=\textwidth]{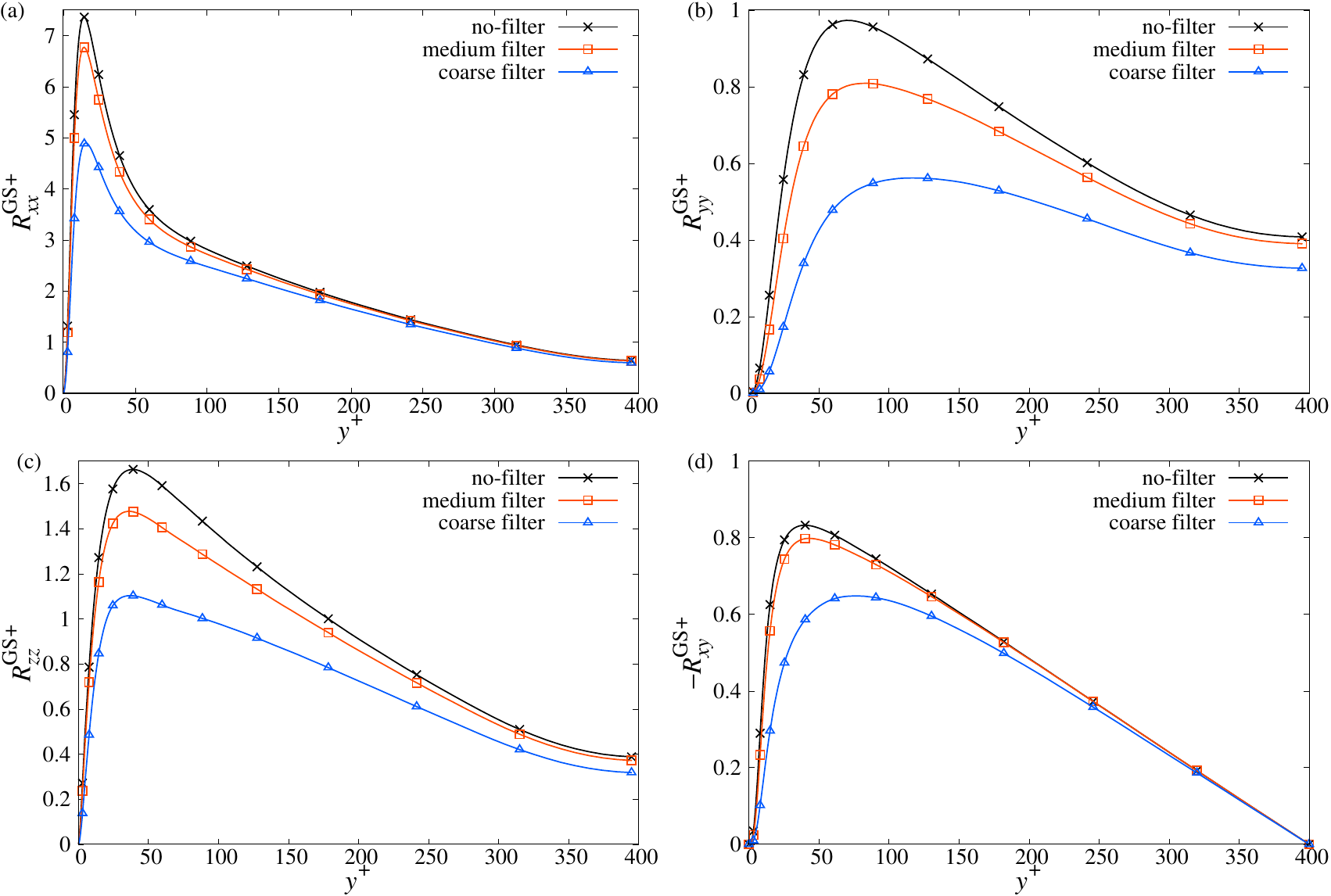}
\caption{Profiles of GS Reynolds stress for (a) streamwise, (b) wall-normal, (c) spanwise, and (d) shear components. The black line with crosses depicts the result of the DNS without filter operation.}
\label{fig:1}
\end{figure}

Figure~\ref{fig:1} shows the profiles of each non-zero component of GS Reynolds stress. The profiles for DNS without filter operation are also plotted. Most of the turbulent velocity fluctuations are resolved in the GS for the medium filter case. The ratio of $K^\mathrm{GS} (=\langle \overline{u}_i' \overline{u}_i' \rangle/2 = R^\mathrm{GS}_{ii}/2)$ to the total turbulent kinetic energy $K (= \langle u_i' u_i' \rangle/2 = K^\mathrm{GS} + \langle \tau^\mathrm{sgs}_{ii} \rangle/2))$ yields approximately greater than 90\% over the entire region $K^\mathrm{GS}/K \gtrsim 0.9$, which is within the tolerance of the conventional LES \cite{popebook}. However, for the coarse filter case, the ratio $K^\mathrm{GS}/K$ is less than 80\% in $y^+ < 100$. Hence, the conventional eddy-viscosity-based LES with a coarse grid fails to predict the statistics of channel flows.

\begin{figure}[tb]
\centering
%
%
\includegraphics[width=0.74\textwidth]{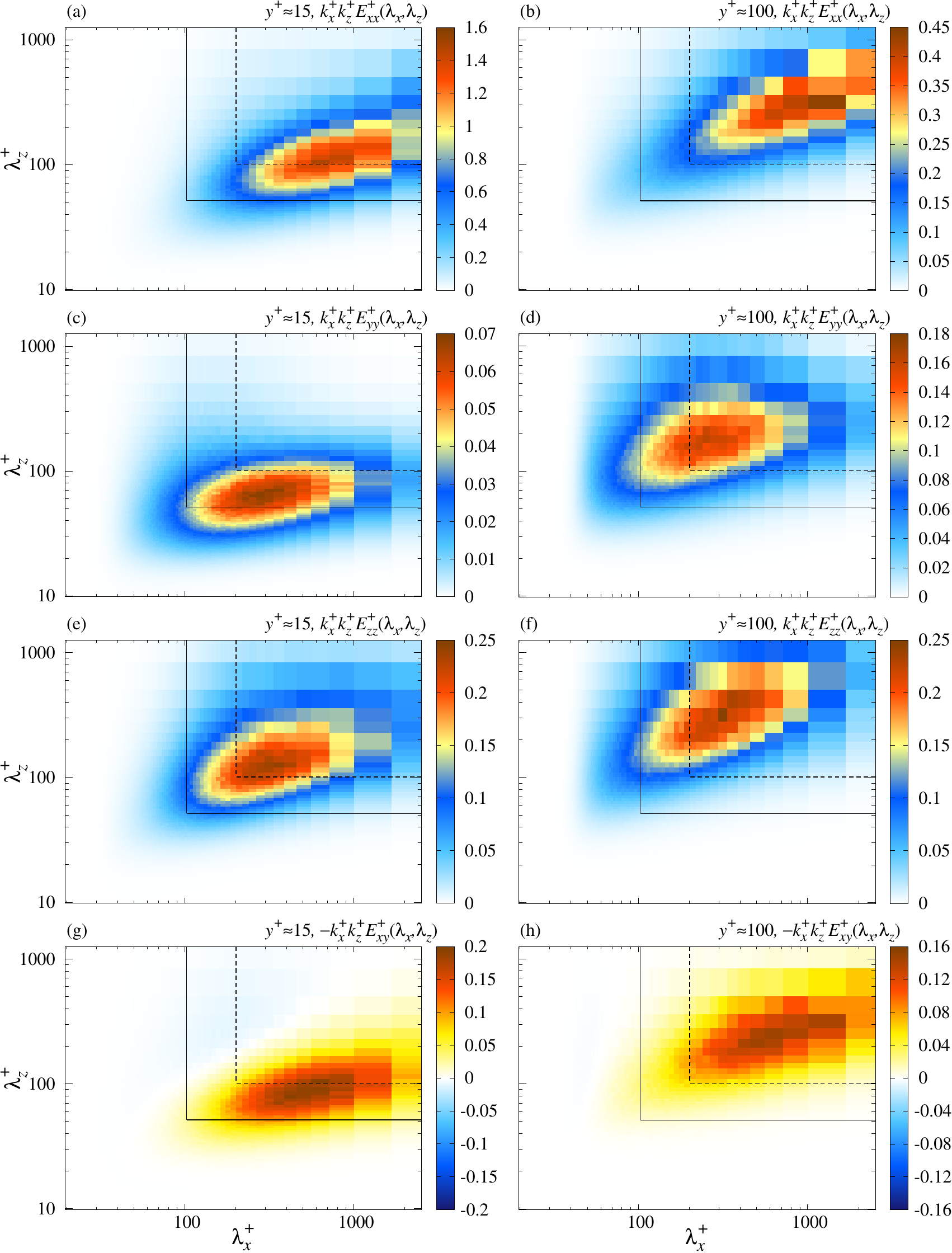}
%
\caption{Pre-multiplied Reynolds stress spectrum $k_x k_z E_{ij} (\lambda_x, \lambda_z)$ where $E_{ij} = \Re \langle \tilde{u}_i' \tilde{u}_j'{}^* \rangle$, which is defined by the unfiltered velocity field. (a,b) Streamwise, (c,d) wall-normal, (e,f) spanwise, and (g,h) shear components. Wall-normal heights for (a,c,e,g) and (b,d,f,h) are $y^+ \approx 15$ and $y^+ \approx 100$, respectively. The solid and dashed lines depict the medium and coarse filter length scales, respectively.}
\label{fig:2}
\end{figure}

To understand the physical meaning of the filter length, two-dimensional pre-multiplied spectra of the Reynolds stress are shown in Fig.~\ref{fig:2}. Here and hereafter, $k_x$ and $k_z$ denote their moduli when considering the statistical values. We chose two heights: the near-wall region $y^+ \approx 15$ and slightly away from the wall $y^+ \approx 100 (y/h \approx 0.25)$. For the medium filter, almost all the components are well-resolved. The wall-normal component is partially filtered out in the near-wall region even for the medium filter. The ratio of wall-normal GS velocity fluctuation to the unfiltered one is less than 80\% in $y^+ < 50$ for the medium filter case. This may be one of the reasons why the wall-normal velocity fluctuation is often underestimated in LESs (see Appendix~\ref{sec:c}). To predict the anisotropy in the near-wall region of shear flows more accurately, a finer grid may be required in the spanwise direction, for example, $\Delta z^+ < 20$. The length scale $\Delta z^+ = 20$ is comparable to that specified by the inverse energy transfer in the spanwise scale proposed by Cimarelli and De Angelis \cite{cda2012}. 

For the coarse filter, half of the peak of the streamwise spectrum is filtered out in the near-wall region $y^+ \approx 15$. This indicates that the filter length scale lies within the energy-containing scale. In addition, most of the wall-normal and shear stress spectra are contained in the SGS. Therefore, conventional isotropic eddy-viscosity models are not valid for coarse filters. Furthermore, in the region slightly away from the wall $y^+ \approx 100$, the wall-normal and spanwise spectra are partially filtered out, whereas the streamwise and shear stress spectra are well-resolved. Hence, the SGS anisotropy is significant even in the region slightly away from the wall in the coarse filter case.

\subsection{\label{sec:level3c}Mean SGS shear stress}

\begin{figure}[tb]
\centering
\includegraphics[width=0.6\textwidth]{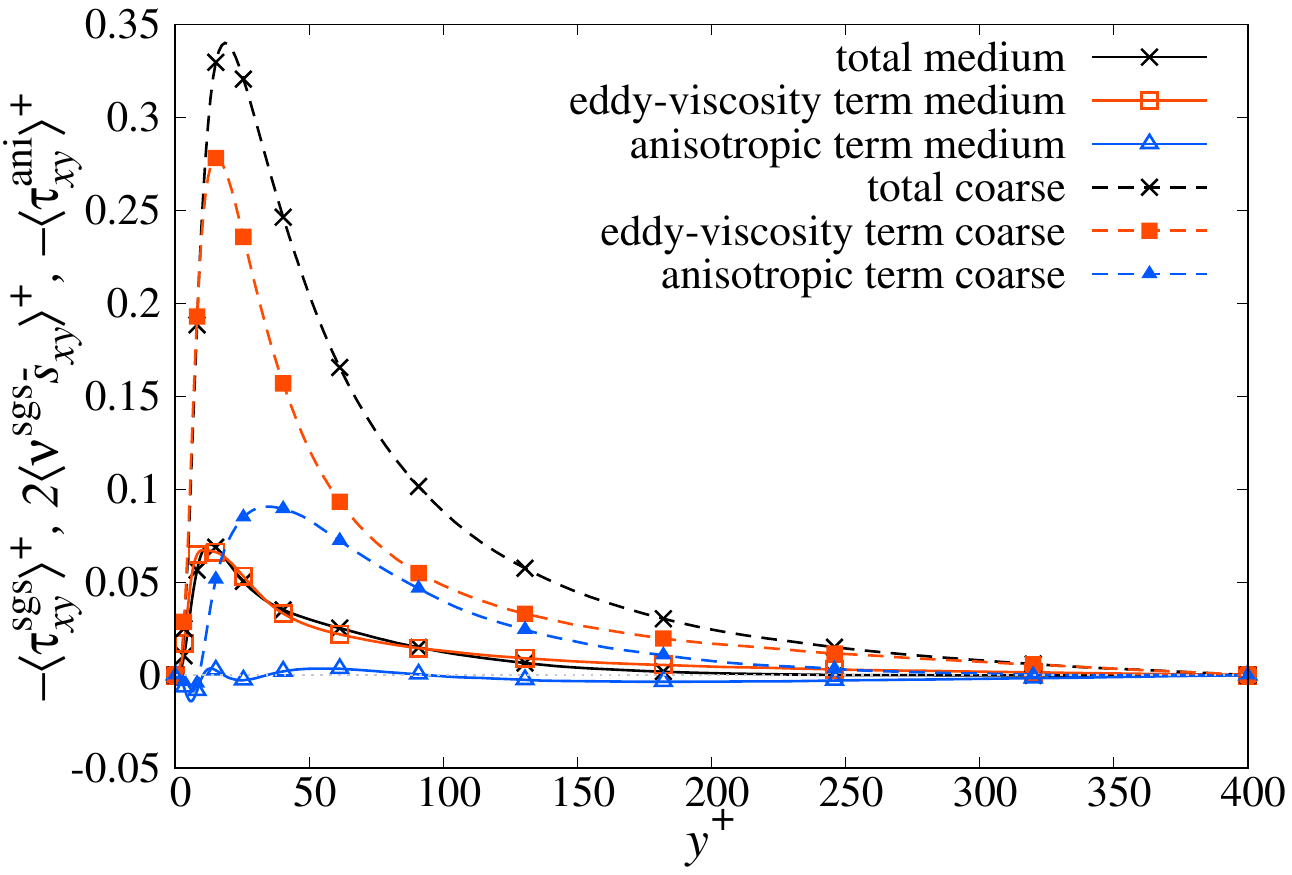}
\caption{Profiles of SGS shear stresses. The solid and dashed lines represent the medium and coarse filter cases, respectively. For both filter sizes, the black, red, and blue lines depict the total SGS stress $-\langle \tau^\mathrm{sgs}_{xy} \rangle$, eddy-viscosity term $2 \langle \nu^\mathrm{sgs} \overline{s}_{xy} \rangle$, and anisotropic term $-\langle \tau^\mathrm{ani}_{xy} \rangle$ given by Eq.~(\ref{eq:5}), respectively. Here and hereafter, the light-gray dotted line is the zero line.}
\label{fig:3}
\end{figure}

It is important to verify the mean SGS shear stress profile. In general, it is difficult to reproduce both the energy transfer rate and mean SGS shear stress using only the eddy-viscosity term \cite{meneveau1994,jm2000,lm2004}. Because the eddy viscosity in this study is determined to reproduce the energy transfer rate, it does not necessarily predict the accurate mean SGS shear stress. Figure~\ref{fig:3} shows the profiles of mean SGS shear stress in terms of the SGS stress decomposition. For the medium filter, the contribution of anisotropic stress is negligible. Therefore, the eddy-viscosity models can predict both the energy transfer rate and mean SGS shear stress for the medium filter case. This result provides an interpretation of why LES with an eddy-viscosity model works well for the medium grid resolution. In contrast, for the coarse filter, anisotropic stress is comparable to the eddy-viscosity term. Hence, the anisotropic stress is necessary for the coarse filter case to predict both the energy transfer rate and mean SGS shear stress.

\subsection{\label{sec:level3d}Budget for GS turbulent kinetic energy}

\begin{figure}[tb]
\centering
 \includegraphics[width=\textwidth]{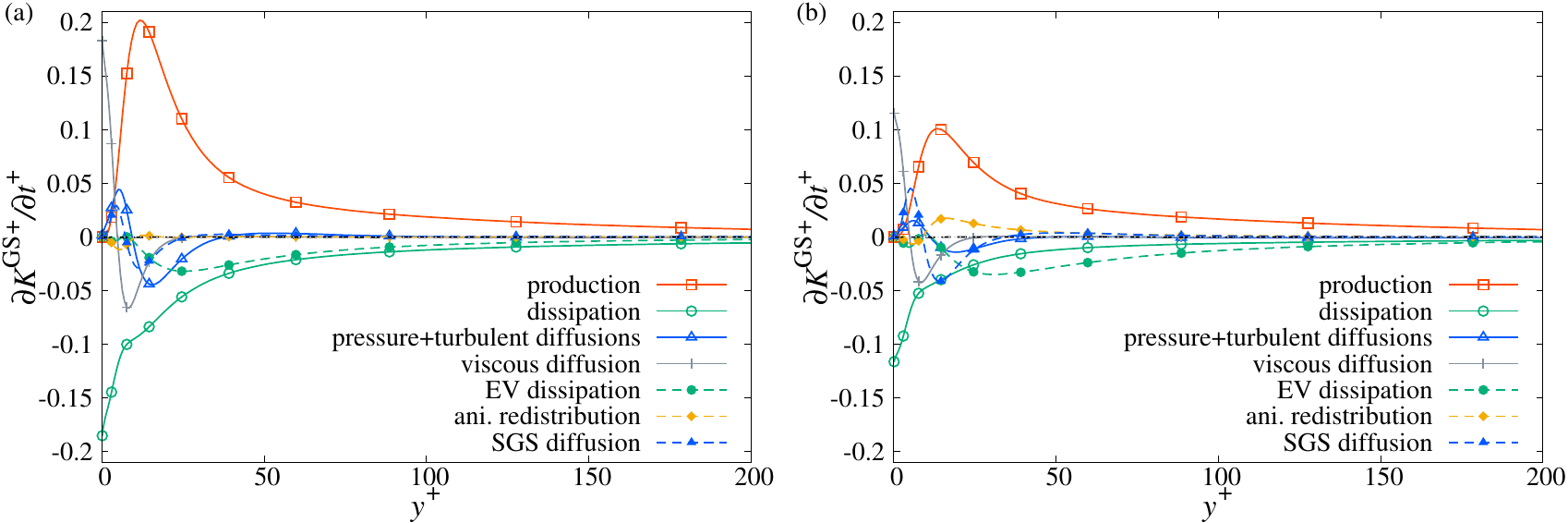}
\caption{Budget for the GS turbulent kinetic energy $K^\mathrm{GS}$ for (a) medium and (b) coarse filter cases. Here and hereafter, the black dashed line depicts the residual or sum of all terms for the budgets.}
\label{fig:4}
\end{figure}

Figure~\ref{fig:4} shows the budget for GS turbulent kinetic energy $K^\mathrm{GS}$. Each term corresponds to half of the trace of Eqs.~(\ref{eq:7a})--(\ref{eq:7f}), (\ref{eq:9a}), (\ref{eq:9b}), and (\ref{eq:7h}). 
Here and hereafter, we only show $y^+ \le 200$ for the budgets to highlight the near-wall features. The trend of each term of the budget is almost monotonic in $y^+ > 200$.
The pressure redistribution (\ref{eq:7c}) is not plotted because it is traceless. For the conventional unfiltered turbulent kinetic energy budget, the production almost balances the dissipation in the region away from the wall $y^+ > 30$ (see e.g. Ref.~\cite{lm2015}). The eddy-viscosity dissipation term also contributes to the budget as an energy sink for the GS turbulent kinetic energy budget. As expected, the contribution of eddy-viscosity dissipation becomes dominant for the coarse filter. The anisotropic redistribution term is negligible for the medium filter. In contrast, this term has a positive value in the near-wall region for the coarse filter. These results are consistent with the profiles of $\langle \tau^\mathrm{ani}_{xy} \rangle$ shown in Fig.~\ref{fig:3}. Because the trace of the anisotropic redistribution is given by Eq.~(\ref{eq:10}), it becomes prominent when the mean anisotropic stress $\langle \tau^\mathrm{ani}_{xy} \rangle$ increases. Note that the SGS dissipation, which is the sum of eddy-viscosity dissipation and anisotropic redistribution, is positive at $y^+ \approx 12$ for the coarse filter. Furthermore, the eddy-viscosity dissipation is also positive at $y^+ \approx 10$, although its value is much smaller than that of SGS dissipation. The productive contribution or averaged backward energy transfer of SGS dissipation in the near-wall region has already been discussed \cite{harteletal1994,domaradzkietal1994,hk1998,cda2011,cda2012}. Several studies suggested a relationship between backward energy transfer and coherent structures observed in the near-wall region of turbulent shear flows \cite{harteletal1994,piomellietal1996,cda2012,hamba2019}. We discuss this in Sec.~\ref{sec:level4a}.

\begin{figure}[tb]
\centering
\includegraphics[width=\textwidth]{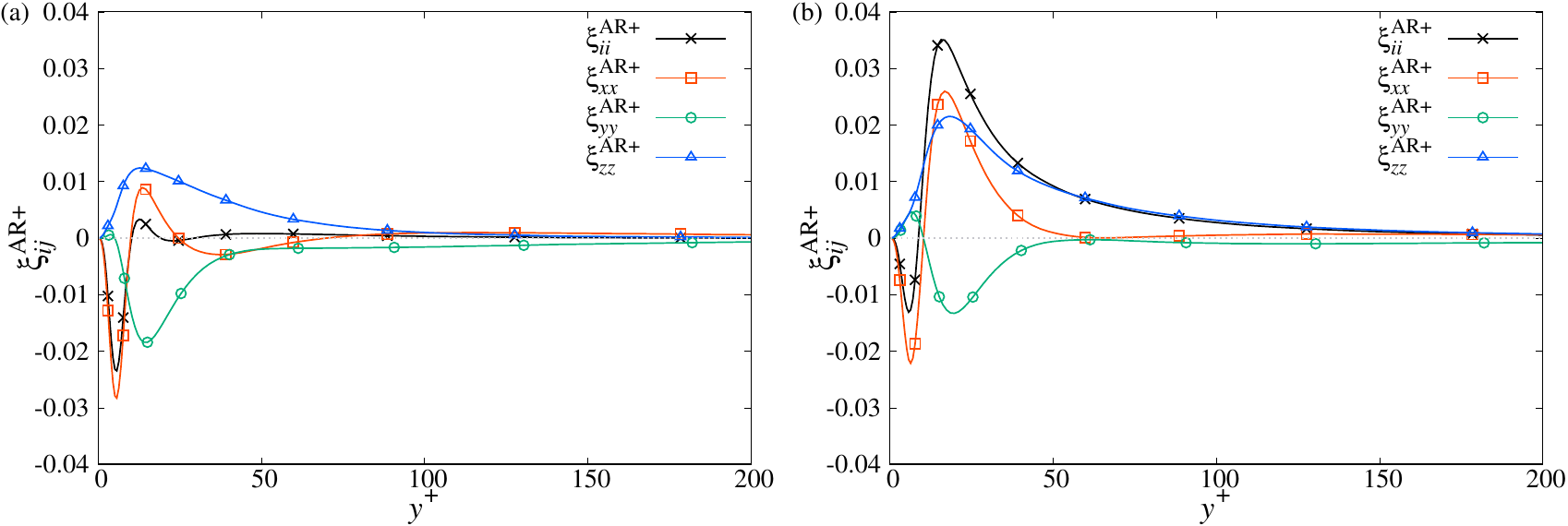}
\caption{Profiles of the normal components of anisotropic redistribution $\xi^\mathrm{GS}_{\alpha \alpha}$ where summation is not taken for $\alpha (= x,y,z)$ for (a) medium and (b) coarse filter cases. The black line depicts the trace part $\xi^\mathrm{AR}_{ii}$.}
\label{fig:5}
\end{figure}

The anisotropic stress is not necessarily negligible even when the trace of the anisotropic redistribution term is small when compared with other terms. Figure~\ref{fig:5} shows the profiles of the normal components of anisotropic redistribution. For the medium filter, the wall-normal and spanwise components are almost canceled out. For the coarse filter, the positive contributions of the streamwise and spanwise components increase at $y^+ \approx 20$. These two components contribute positively to the GS turbulent kinetic energy budget in the near-wall region as shown in Fig.~\ref{fig:4}(b). In addition, the spanwise component is dominant in $50 < y^+ < 100$ for the coarse filter. Notably, the spanwise component of the anisotropic redistribution is always positive for both filter sizes. In turbulent channel flows, the production term is zero in the budget for the spanwise component of GS Reynolds stress $R^\mathrm{GS}_{zz}$; that is, $P^\mathrm{GS}_{zz} = 0$. The productive term, which is the pressure redistribution in the conventional unfiltered budget, plays a significant role in the budget for spanwise GS Reynolds stress. In the next subsection, we examine the budget for the normal components of GS Reynolds stress.

\subsection{\label{sec:level3e}Budget for GS Reynolds stress}

In this study, we focus on the budget only for the normal components of GS Reynolds stress. A detailed analysis of the contributions of anisotropic SGS stress to the shear stress budget was already demonstrated by Abe \cite{abe2019}, which concluded that anisotropic SGS stress is essential to reproduce the productive contribution to the shear stress budget.

\subsubsection{\label{sec:level3e1}Streamwise component}

\begin{figure}[tb]
\centering
\includegraphics[width=\textwidth]{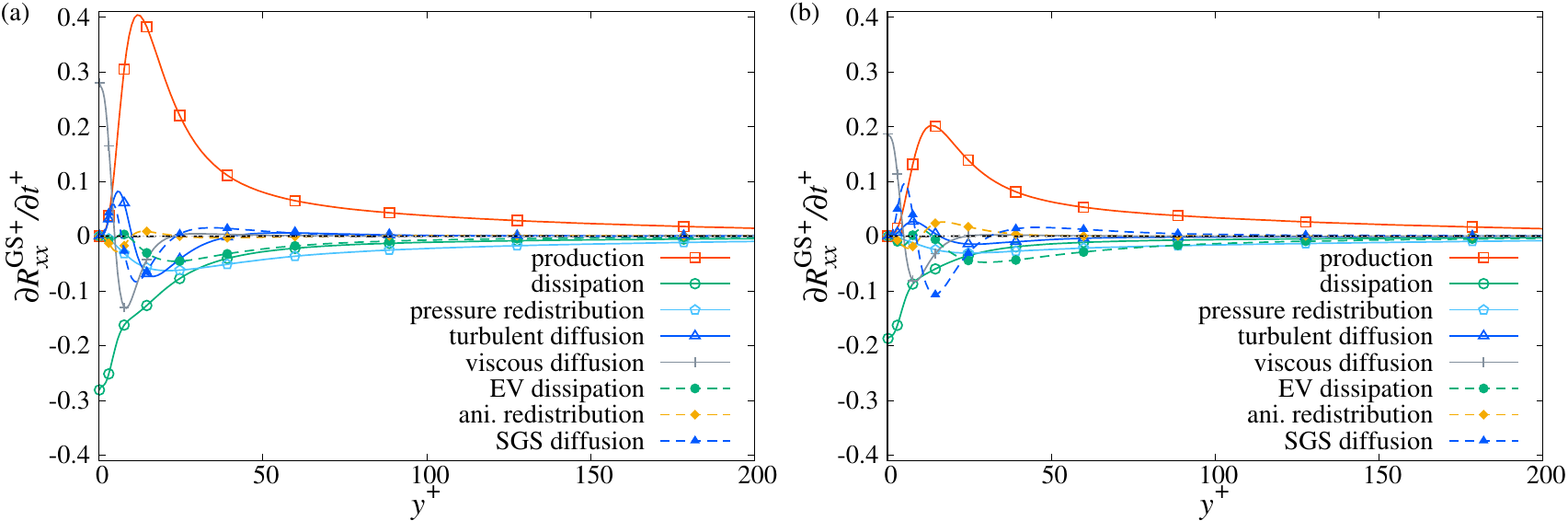}
\caption{Budget for the streamwise component of GS Reynolds stress $R^\mathrm{GS}_{xx}$ for (a) medium and (b) coarse filter cases.}
\label{fig:6}
\end{figure}

Figure~\ref{fig:6} shows the budget for the streamwise component of GS Reynolds stress. The basic profiles of each term are almost the same as those of the GS turbulent kinetic energy budget shown in Fig.~\ref{fig:4}. The pressure redistribution term, which transfers the energy from the streamwise component to the other two components, is an exception. The intensity of the pressure redistribution is small when compared to the eddy-viscosity dissipation for the coarse filter. However, this does not imply that the redistribution among the normal components is negligible because pressure redistribution is an essential source term in the wall-normal and spanwise components.

\subsubsection{\label{sec:level3e2}Wall-normal component}

\begin{figure}[tb]
\centering
\includegraphics[width=\textwidth]{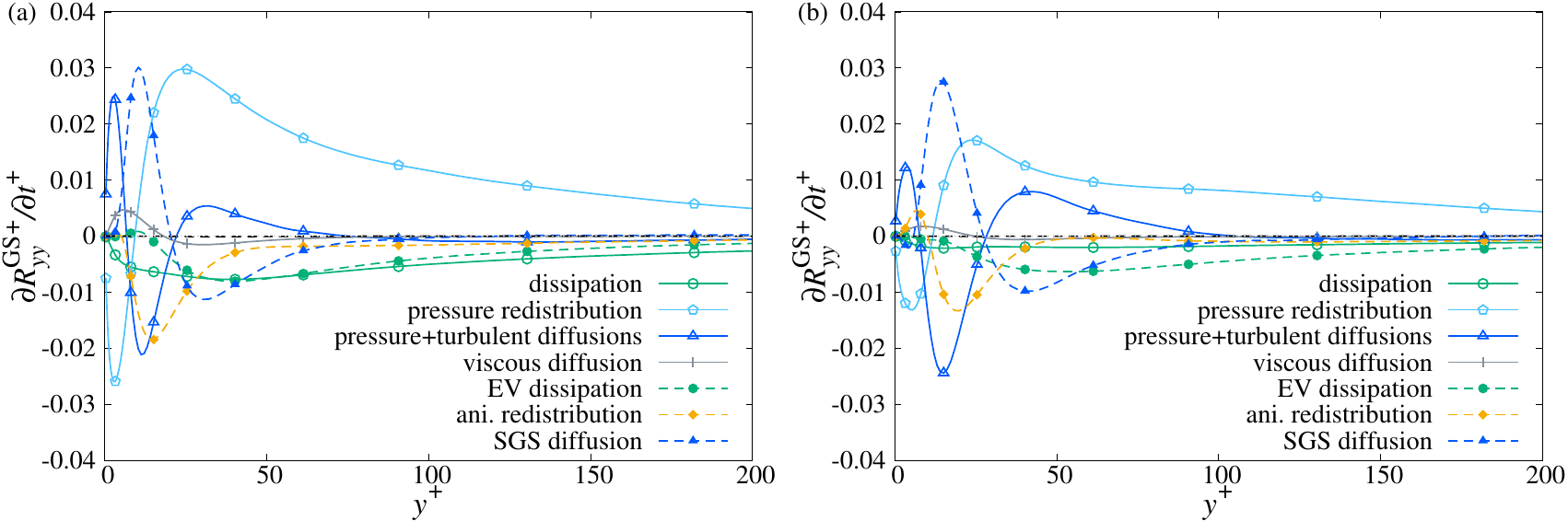}
\caption{Budget for the wall-normal component of GS Reynolds stress $R^\mathrm{GS}_{yy}$ for (a) medium and (b) coarse filter cases.}
\label{fig:7}
\end{figure}

Figure~\ref{fig:7} shows the budget for the wall-normal component of GS Reynolds stress. The pressure redistribution has a large productive contribution except for the close vicinity of the wall $y^+ <10$. The viscous and eddy-viscosity dissipations are the leading terms of the negative contribution for both filter sizes. For the coarse filter, the eddy-viscosity dissipation is dominant, as seen in the budget for the streamwise component. In the near-wall to buffer region $y^+ < 100$, the SGS diffusion term also contributes significantly to the budget and plays a key role in counterbalancing the sum of the pressure and turbulent diffusions. The details of SGS diffusion are provided in Appendix~\ref{sec:a}. The anisotropic redistribution has a large negative contribution compared to the viscous and eddy-viscosity dissipations in the near-wall region $y^+ < 50$. Conventional eddy-viscosity models may account for this negative contribution in the \textit{a posteriori} tests, although the intensity could be small. Thus, the leading productive and dissipative contributions to the wall-normal GS Reynolds stress can be reproduced by the pressure redistribution, viscous dissipation, and eddy-viscosity dissipation terms.

\subsubsection{\label{sec:level3e3}Spanwise component}

\begin{figure}[tb]
\centering
\includegraphics[width=\textwidth]{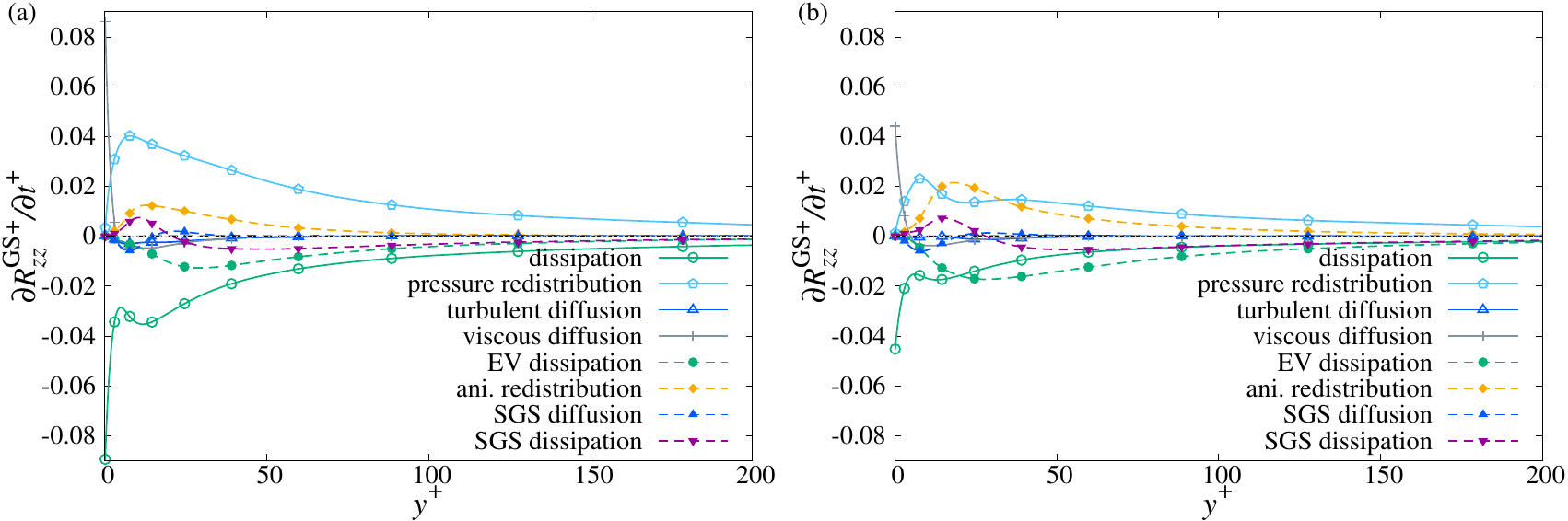}
\caption{Budget for the spanwise component of GS Reynolds stress $R^\mathrm{GS}_{zz}$ for (a) medium and (b) coarse filter cases. We also plot the sum of eddy-viscosity dissipation and anisotropic redistribution, which is depicted as SGS dissipation in the purple dashed line with inverted triangles.}
\label{fig:8}
\end{figure}

Figure~\ref{fig:8} shows the budget for the spanwise component of GS Reynolds stress. The pressure redistribution has a leading productive contribution similar to the wall-normal component shown in Fig.~\ref{fig:7}. In addition, the viscous and eddy-viscosity dissipations are also the leading terms of the negative contribution. An important finding is that the anisotropic redistribution is always positive in the spanwise GS Reynolds stress budget as shown in Fig.~\ref{fig:5}. Furthermore, the sum of eddy-viscosity dissipation and anisotropic redistribution is positive near the wall $y^+ \approx 20$ for both filter sizes. The intensity of the anisotropic redistribution is relatively small compared to that of the pressure redistribution for the medium filter. In contrast, the anisotropic redistribution is comparable to the pressure redistribution in the near-wall to buffer region $y^+ < 100$ for the coarse filter. Therefore, this positive contribution of anisotropic redistribution is indispensable to the generation mechanism of the spanwise velocity fluctuation in the GS or the resolved scale for the coarse filter case. Even for the medium filter, we infer that the lack of anisotropic redistribution will lead to an underestimation of the GS spanwise velocity fluctuation (see also Appendix~\ref{sec:c}). The underestimation of GS spanwise velocity fluctuation may alter the structure of the wall-bounded turbulent shear flows, for example, coherent structures in the near-wall region. Hamba \cite{hamba2019} demonstrated that the conditional averaged velocity field regarding the inverse cascade of the spanwise velocity fluctuation represents the streamwise elongated vorticity structure, which represents the coherent structure in wall-bounded turbulent shear flows. A relationship between backward scatter in terms of kinetic energy and coherent structures has also been suggested \cite{harteletal1994,piomellietal1996,cda2012}. The present analysis suggests that the anisotropic SGS stress reproducing the productive contribution to the spanwise GS Reynolds stress budget is key to improving SGS models.

\section{\label{sec:level4}Discussion}

In Sec.~\ref{sec:level3e}, we demonstrated that the anisotropic redistribution term contributes positively to the streamwise and spanwise components of GS Reynolds stress. Conventional eddy-viscosity models cannot represent the productive contribution. In this section, first, we discuss the physical interpretation of the productive contribution in terms of the budget equation for GS Reynolds stress spectrum. Second, we perform an \textit{a priori} test of the anisotropic redistribution term based on several existing model expressions.

\subsection{\label{sec:level4a}Budget for GS Reynolds stress spectrum}

Several studies have discussed the relationship between inverse cascade and coherent structures in wall-bounded turbulent shear flows \cite{harteletal1994,piomellietal1996,cda2012,hamba2019}. Spectral analysis is a fundamental tool used to study the effect of each term in the budget on structures represented by specific scales \cite{ka2018,lm2019}. A representative of the near-wall structure is streaky structures whose spanwise spacing or wavelength is $\lambda_z^+ \sim 100$ \cite{klineetal1967,jm1991}. According to the self-sustaining process of wall-bounded turbulent shear flows, streaky structures are generated by streamwise vortices, and the breakdown of the streaks generates the source modes of the nonlinear interaction that generates the streamwise vortices \cite{hamiltonetal1995,waleffe1997}. In particular, the generation of streamwise vortices does not agree with the dissipative property of eddy viscosity \cite{hamba2019}. To determine the relationship between these processes and the productive contribution of anisotropic redistribution in the streamwise and spanwise components of GS Reynolds stress, we examine the budget for $E^\mathrm{GS}_{xx} (k_x)$ and $E^\mathrm{GS}_{zz} (k_z)$. Here and hereafter, we simply denote $E^\mathrm{GS}_{ij} (k_x) = \sum_{n_z} E^\mathrm{GS}_{ij} (k_x,k_z)$ or $E^\mathrm{GS}_{ij} (k_z) = \sum_{n_x} E^\mathrm{GS}_{ij} (k_x,k_z)$. Although $E^\mathrm{GS}_{zz} (k_z)$ itself does not directly represent the vorticity, the spanwise velocity fluctuation accompanied by the nonzero spanwise wavenumber is related to the streamwise vortices as $\tilde{\omega}_x (k_z) = \partial \tilde{u}_z (k_z)/\partial y - \mathrm{i} k_z \tilde{u}_y (k_z)$.

\subsubsection{\label{sec:level4a1}Streamwise component}

\begin{figure}[tb]
\centering
\includegraphics[width=\textwidth]{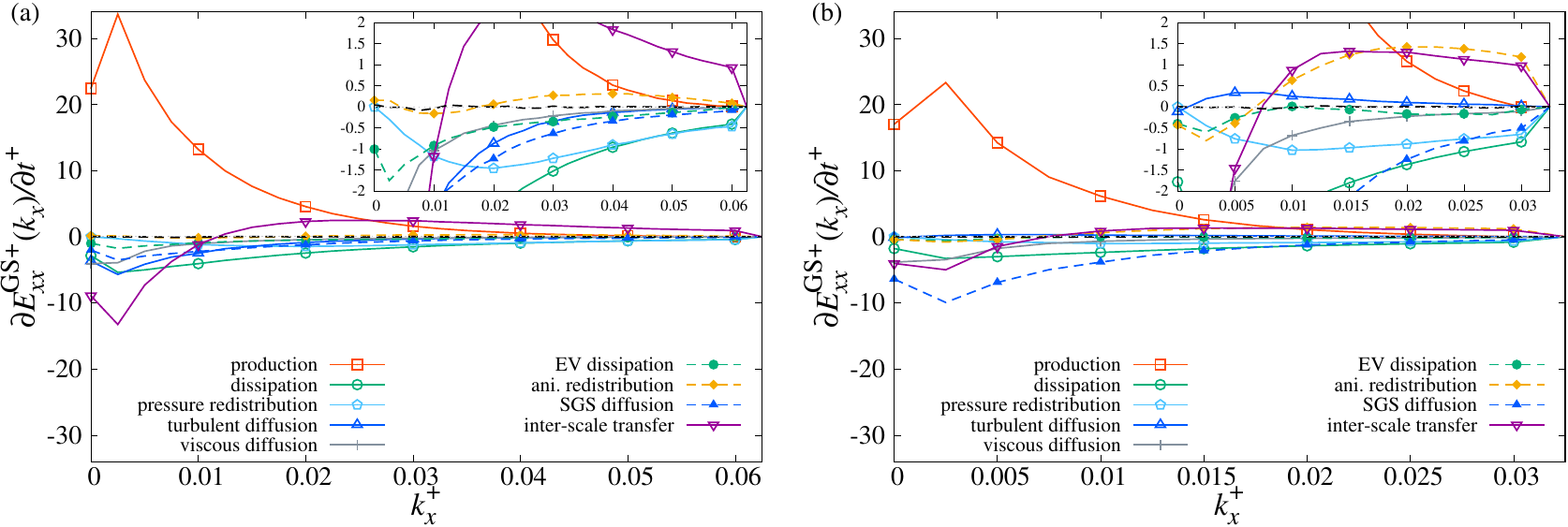}
\caption{Budget for the streamwise component of GS Reynolds stress spectrum in the streamwise wavenumber space $E^\mathrm{GS}_{xx}(k_x)$ for (a) medium and (b) coarse filter cases at $y^+ \approx 15$. The insets show the small vertical axis range to focus on the small-scale or high-wavenumber region.}
\label{fig:9}
\end{figure}

Figure~\ref{fig:9} shows the budget for the streamwise component of GS Reynolds stress spectrum in the streamwise wavenumber space $E^\mathrm{GS}_{xx} (k_x)$ at $y^+\approx 15$. We plot the budget with a linear scale in $k_x$ instead of a log scale or wavelength scale to depict the $k_x=0$ mode. For both filter sizes, the production term contributes significantly in the low-wavenumber region. For the medium filter, the gain by the interscale interaction is balanced with the loss by viscous dissipation and pressure redistribution in the high-wavenumber region. In contrast, for the coarse filter, the anisotropic redistribution has a positive contribution in the high-wavenumber region $k_x^+ > 0.01$ ($\lambda_x^+ \lesssim 600$). The contribution of anisotropic redistribution to the high-wavenumber mode was already highlighted by Inagaki and Kobayashi \cite{ik2020} in the low-Reynolds number case. They suggested that in coarse LES using only the eddy-viscosity term, the absence of enhancement of small scales causes a longer streamwise velocity correlation in the $x$ direction than that of filtered DNS. In other words, the anisotropic term other than eddy viscosity is needed in coarse LES to predict the streamwise velocity correlation in the $x$ direction comparable to that of filtered DNS.

\begin{figure}[tb]
\centering
\includegraphics[width=\textwidth]{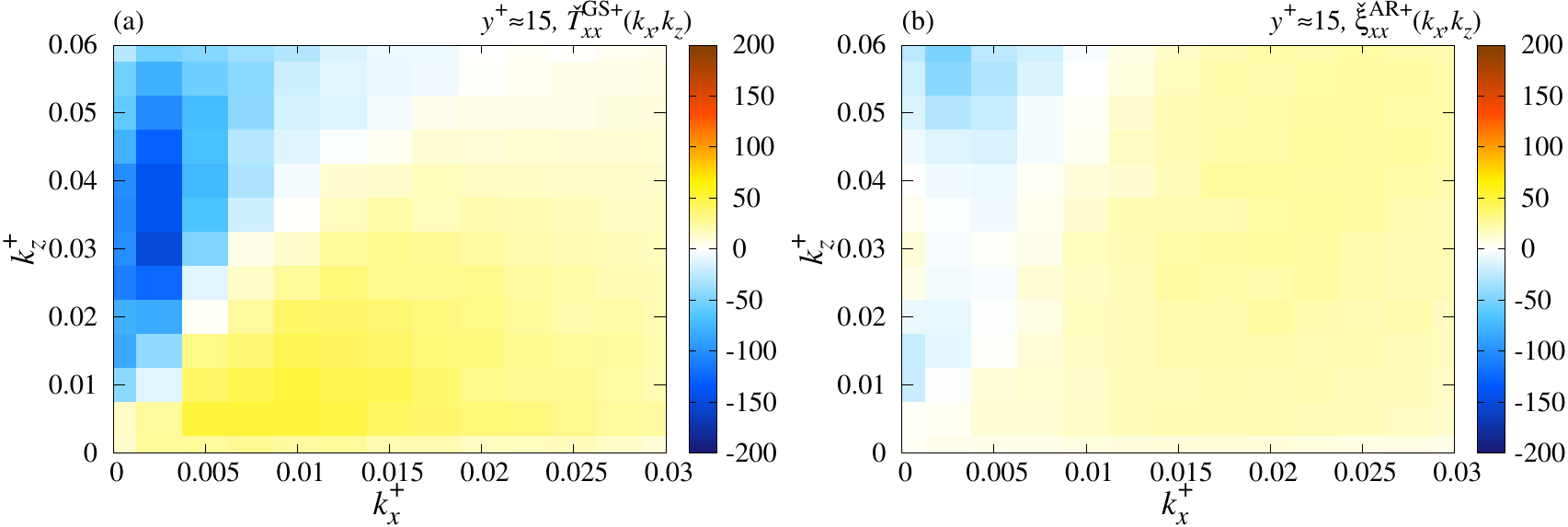}
\caption{Two-dimensional spectra of (a) the interscale transfer $\check{T}^\mathrm{GS}_{xx}(k_x,k_z)$ and (b) anisotropic redistribution $\check{\xi}^\mathrm{AR}_{xx} (k_x, k_z)$ for the coarse filter case at $y^+ \approx 15$.}
\label{fig:10}
\end{figure}

To see the details of the gain in the small scales, we examine the two-dimensional spectra of the interscale transfer and anisotropic redistribution in the $k_x$-$k_z$ plane for the coarse filter case. Figure~\ref{fig:10} shows the contributions of interscale transfer and anisotropic redistribution at $y^+ \approx 15$. The interscale transfer term transports $E^\mathrm{GS}_{xx} (k_x,k_z)$ from the region $k_x^+ < 0.01$ and $0.02 < k_z^+ < 0.06$ ($\lambda_x^+ \gtrsim 600$ and $100 \lesssim \lambda_z^+ \lesssim 300$) 
to the modes with a relatively large spanwise length scale where $k_z^+ < 0.03$ ($\lambda_z^+ \gtrsim 200$). In contrast, the anisotropic redistribution term contributes to the high-wavenumber region $k_x^+ > 0.015$ ($\lambda_x^+ \lesssim 400$). 
Therefore, we can interpret that the anisotropic redistribution plays a key role in the amplification of small-scale mode in the streamwise scale. The typical spanwise length scale of streaky structures lies on a scale close to the cutoff $\lambda_z^+ \sim 100 \sim \lambda_z^{\mathrm{c}+}$ for the coarse filter. Focusing on $k_x^+ > 0.015$ and $k_z^+ > 0.04$ ($\lambda_x^+ \lesssim 400$ and $\lambda_z^+ \lesssim 160$), the anisotropic redistribution has a slightly larger contribution to the budget than interscale transfer. This generation of a small streamwise length scale mode with $\lambda_z^+ \sim 100$ is consistent with the streak breakdown. 
We infer that this positive contribution of anisotropic redistribution can improve the performance of LES in the way that it reproduces the streak breakdown-like contribution.

\subsubsection{\label{sec:level4a2}Spanwise component}

\begin{figure}[tb]
\centering
\includegraphics[width=\textwidth]{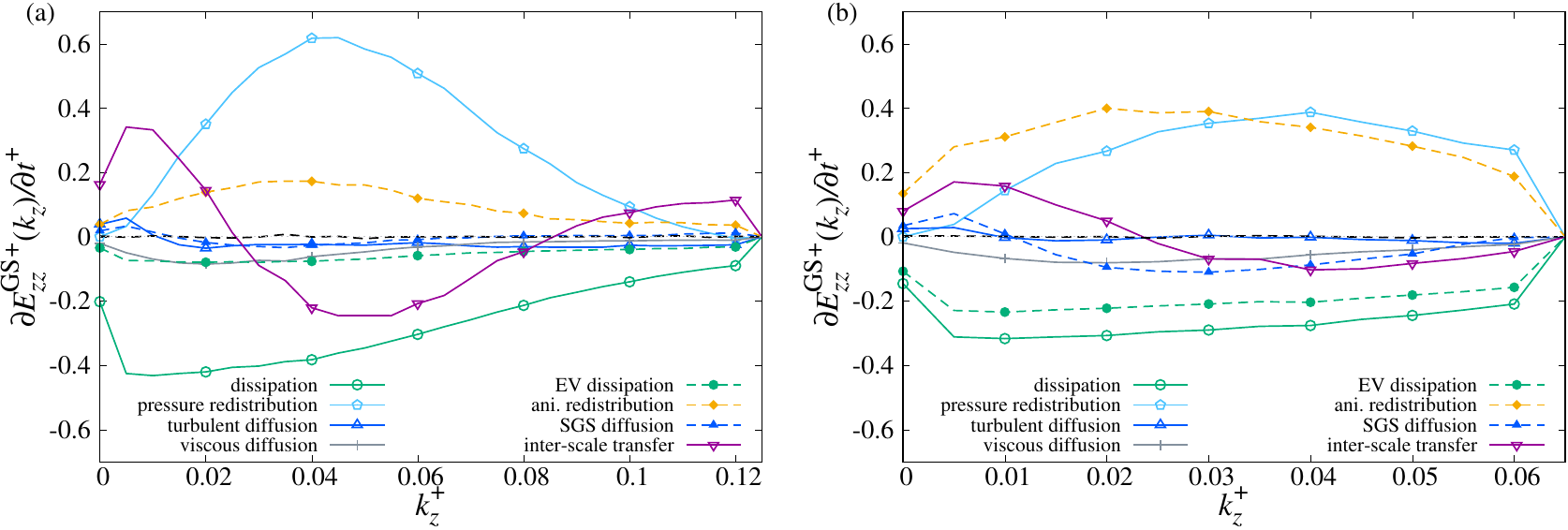}
\caption{Budget for the spanwise component of GS Reynolds stress spectrum in the spanwise wavenumber space $E^\mathrm{GS}_{zz}(k_z)$ for (a) medium and (b) coarse filter cases at $y^+ \approx 15$.}
\label{fig:11}
\end{figure}


Figure~\ref{fig:11} shows the budget for the spanwise component of GS Reynolds stress spectrum in the spanwise wavenumber space $E^\mathrm{GS}_{zz} (k_z)$ at $y^+\approx 15$. The interscale transfer is negative at $k_z^+ = 0.04$ and positive in $k_z^+ < 0.02$ for both filter sizes, which represents the inverse transfer of $E^\mathrm{GS}_{zz} (k_z)$ in the spanwise scale. The inverse transfer of the spanwise velocity fluctuation has already been identified by Hamba \cite{hamba2019}, although it has been demonstrated in the streamwise scale space. In addition, the inverse transfer of kinetic energy in the spanwise scale has already been discussed \cite{cda2012,hamba2018}. For the medium filter, the interscale transfer changes the sign at $k_z^+ \approx 0.08$ ($\lambda_z^+ \approx 80$) in the high-wavenumber region. This critical length scale is larger than that observed in the analysis based on the Kolmogorov equation \cite{cda2012} or the scale energy density in terms of a filter function \cite{hamba2018}. They suggested that the critical length is $r_z^+ \approx 20$ where $r_z$ denotes the distance between the two velocity fields composing the scale. Note that the wavelength should be twice the distance between the two velocity fields; that is $\lambda_z = 2r_z$. The shift in the critical length of interscale transfer is caused by the absence of small scales, owing to the filtering operation. Nevertheless, we can infer that a large amount of the interscale interaction in the budget of $E^\mathrm{GS}_{zz} (k_z)$ including inverse transfer is resolved in the medium filter case. In contrast, for the coarse filter, the forward cascade of $E^\mathrm{GS}_{zz} (k_z)$ is completely unresolved. The unresolved interscale interaction should be converted to the $\tau^\mathrm{sgs}_{\ell \ell}$-related part of the pressure redistribution, eddy-viscosity dissipation, and anisotropic redistribution. Part of the forward cascade may convert to eddy-viscosity dissipation. However, for both filter sizes, the positive contribution of anisotropic redistribution is larger than the loss via eddy-viscosity dissipation in the entire wavenumber. In addition, the anisotropic redistribution is comparable to the pressure redistribution for the coarse filter.

\begin{figure}[tb]
\centering
\includegraphics[width=\textwidth]{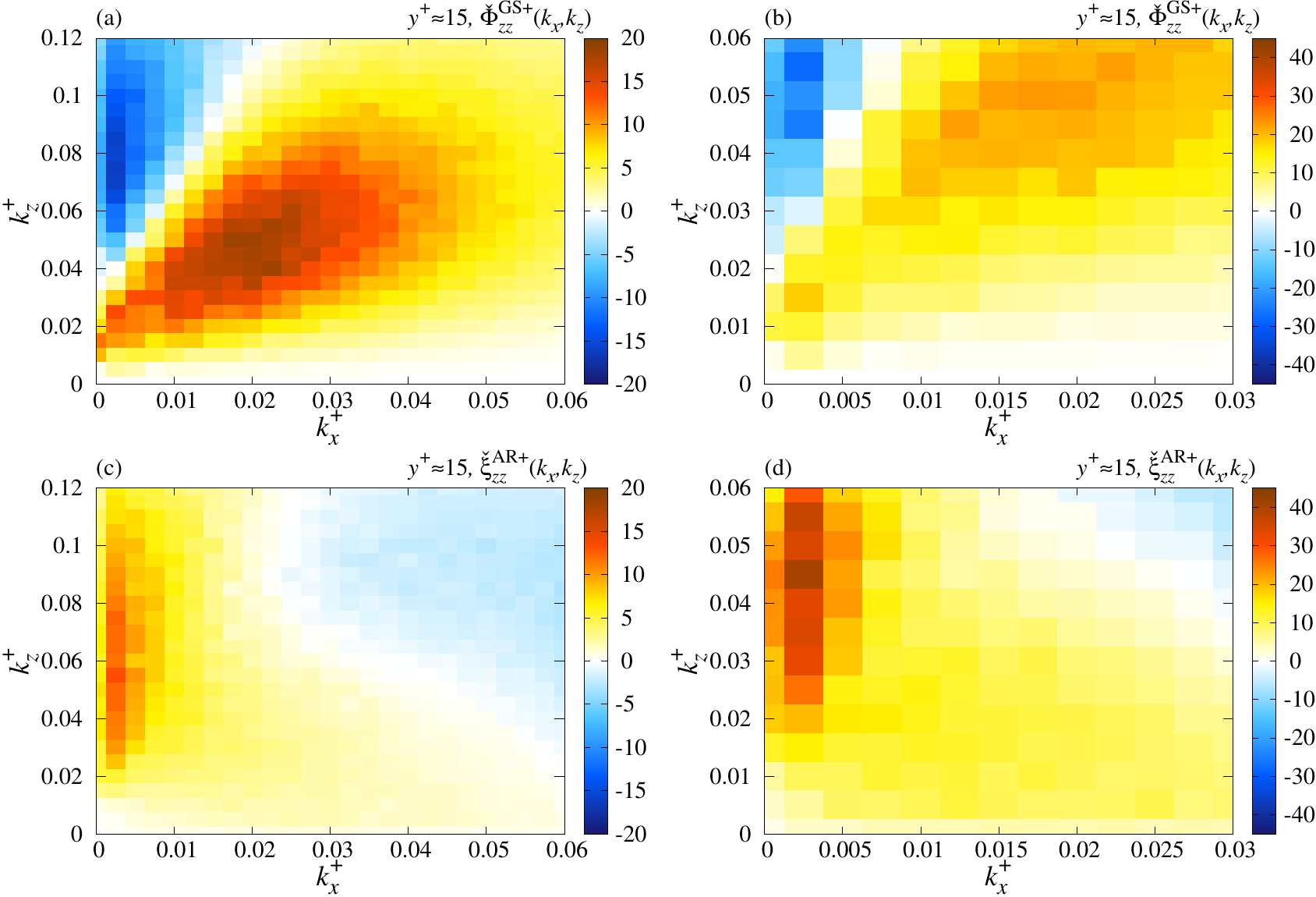}
\caption{Two-dimensional spectra of (a,b) the pressure redistribution $\check{\Phi}^\mathrm{GS}_{zz}(k_x,k_z)$ and (c,d) anisotropic redistribution $\check{\xi}^\mathrm{AR}_{zz} (k_x, k_z)$ for (a,c) medium and (b,d) coarse filter cases at $y^+ \approx 15$.}
\label{fig:12}
\end{figure}


To observe the relationship between the coherent structures and positive contributions to the budget, we examine the two-dimensional spectra of the pressure and anisotropic redistributions in the $k_x$-$k_z$ plane. Figure~\ref{fig:12} shows the two-dimensional spectra of the pressure and anisotropic redistributions at $y^+ \approx 15$. The pressure redistribution is negative in the region $k_x^+ < 0.005$ and $k_z^+ > 0.04$ ($\lambda_x^+ \gtrsim 1200$ and $\lambda_z^+ \lesssim 160$) for both filter sizes. This region is consistent with the scale of streamwise vortices in the self-sustaining process \cite{hamiltonetal1995,waleffe1997}. Because $E^\mathrm{GS}_{zz} (k_z)$ can be related to the streamwise vortices, we infer that the pressure redistribution attenuates the streamwise vortices. The anisotropic redistribution is positive in this region and peaks at $k_z^+ \approx 0.05$ ($\lambda_z^+ \approx 130$), which is consistent with the typical spanwise spacing of streaks $\lambda_z^+ \sim 100$. Therefore, we infer that the positive contribution of anisotropic redistribution to the spanwise GS Reynolds stress spectrum is related to the generation of streamwise vortices in the self-sustaining process \cite{hamiltonetal1995,waleffe1997}. In other words, the generation mechanism of streamwise vortices through the SGS stress vanishes if we employ only the eddy-viscosity model in the LES. 
As we can see from Fig.~\ref{fig:12}(c), the anisotropic redistribution has a large value at a low-$k_x$ region, whereas it is large even at a relatively high-$k_z$ region. Therefore, we can infer that the filter in the $z$ direction is more critical than that in the $x$ direction for the anisotropic redistribution.
The anisotropic redistribution is a significant source term at this scale even in the medium filter case. Hence, by employing a proper anisotropic SGS stress reproducing the productive contribution to the spanwise GS Reynolds stress, the prediction of the statistics in LES for the coarse-to-medium filter cases can be improved.

\subsection{\label{sec:level4b}\textit{A priori} test of anisotropic redistribution term}

A classical idea for implementing backward scatter in LES is stochastic modeling \cite{leith1990}. Langford and Moser \cite{lm1999} demonstrated that the force from SGS stress is mostly stochastic in their analysis of homogeneous isotropic turbulence. They suggested that the SGS model may only be able to estimate the average energy transfer rate from GS to SGS because of its stochastic nature. However, the stochastic approach cannot predict the productive contribution to the GS Reynolds stress budget; that is, even if we add the stochastic forcing term $a_i$ to the filtered Navier-Stokes equations (\ref{eq:2}), its contribution to the budget is always zero $\langle \overline{u}_i' a_j \rangle = 0$ because of its stochastic nature. Therefore, a deterministic model must be employed to reproduce the positive contribution to the GS Reynolds stress budget.

We examine the performance of several model expressions for the anisotropic SGS stress in terms of an \textit{a priori} test of the anisotropic redistribution. In this study, the anisotropic stress is assumed to have no contribution to the energy transfer between GS and SGS, according to Abe \cite{abe2013,abe2019}. Similarly, we examine fundamental models based on the strategy proposed by Abe \cite{abe2013}. Namely, we adopt the following two models for anisotropic stress:
\begin{align}
\tau^\mathrm{ani}_{ij} & = \tau^\mathrm{sgs}_{\ell \ell} \frac{\tau^\mathrm{a}_{ij}|_\mathrm{tl} + 2 \nu^\mathrm{a} \overline{s}_{ij}}{ \tau^\mathrm{a}_{mm} }, \ \ 
\tau^\mathrm{a}_{ij} = (\overline{u}_i - \widehat{\overline{u}}_i) (\overline{u}_j - \widehat{\overline{u}}_j), 
\label{eq:19} \\
\tau^\mathrm{ani}_{ij} & = \tau^\mathrm{sgs}_{\ell \ell} \frac{\tau^\mathrm{a}_{ij}|_\mathrm{tl} + 2 \nu^\mathrm{a} \overline{s}_{ij}}{ \tau^\mathrm{a}_{mm} }, \ \ 
\tau^\mathrm{a}_{ij} = \sum_{\alpha = 1,2,3} \overline{\Delta}_\alpha^2 \frac{\partial \overline{u}_i}{\partial x_\alpha} \frac{\partial \overline{u}_j}{\partial x_\alpha},
\label{eq:20} 
\end{align}
where $\nu^\mathrm{a} = -\tau^\mathrm{a}_{ij} \overline{s}_{ij}/(2 \overline{s}^2)$ for both models, which is introduced to remove energy transfer through $\tau^\mathrm{a}_{ij}$. The first model (\ref{eq:19}) is the scale-similarity model for the SGS Reynolds term $\overline{(u_i-\overline{u}_i )(u_j-\overline{u}_j)} \simeq (\overline{u}_i - \overline{\overline{u}}_i) (\overline{u}_j - \overline{\overline{u}}_j)$ \cite{bardinaetal1983}, although the repeated filter is replaced with the test filter denoted by $\widehat{\cdot}$. The filter length for the test filter is set to twice that of the filter $\overline{\cdot}$; namely, $\widehat{\overline{\Delta}}/\overline{\Delta} = 2$. The model (\ref{eq:19}) is employed in the stabilized mixed model \cite{abe2013,ia2017} with the modeled transport equation of the SGS kinetic energy $\tau^\mathrm{sgs}_{\ell \ell}/2$. The second model (\ref{eq:20}) is the Clark term, which is the leading term in the Taylor expansion of the sum of Leonard and cross terms \cite{clarketal1979}; that is $\overline{\overline{u}_i \overline{u}_j} - \overline{u}_i \overline{u}_j + \overline{\overline{u}_i (u_j - \overline{u}_j)} + \overline{(u_i- \overline{u}_i) \overline{u}_j} \simeq \sum_\alpha \overline{\Delta}_\alpha^2 (\partial \overline{u}_i/\partial x_\alpha) (\partial \overline{u}_j/\partial x_\alpha)/12 + O (\overline{\Delta}^4)$. Inagaki and Kobayashi \cite{ik2020} examined these two models in an \textit{a posteriori} test of turbulent channel flows. They found that the first model provides a better result than the second because small-scale velocity fluctuation is significantly enhanced in the first one.

In addition to these two models, we also examine the following quadratic nonlinear model based on the velocity gradient according to the explicit algebraic SGS stress model \cite{marstorpetal2009,montecchiaetal2017}:
\begin{align}
\tau^\mathrm{ani}_{ij} = - \tau^\mathrm{sgs}_{\ell \ell} \frac{\overline{s}_{im} \overline{w}_{mj} + \overline{s}_{jm} \overline{w}_{mi}}{\overline{s}^2}, \ \ 
\overline{w}_{ij} = \frac{1}{2} \left( \frac{\partial \overline{u}_i}{\partial x_j} - \frac{\partial \overline{u}_j}{\partial x_i} \right).
\label{eq:21}
\end{align}
This model does not contribute to the energy transfer without any artificial treatments because $(\overline{s}_{im} \overline{w}_{mj} + \overline{s}_{jm} \overline{w}_{mi}) \overline{s}_{ij} = 0$. We set the arbitrary numerical coefficient to unity.

We can examine other models, including other normalizations, such as the linear Clark model where the coefficient $\tau^\mathrm{sgs}_{\ell \ell}/\tau^\mathrm{a}_{mm}$ is replaced with a constant or some other nondimensional function. In addition, to remove the energy transfer by the anisotropic term, we may be able to construct treatments other than $\nu^\mathrm{a} \overline{s}_{ij}$. In this study, we restrict ourselves our examination to the existing model expressions. 

It is worth examining the profiles of the anisotropic redistribution term when employing the commonly used eddy viscosities in LES because the definition of eddy viscosity given by Eq.~(\ref{eq:4}) is an ideal one. Then, we also examine the anisotropic redistribution term by using the eddy viscosity based on the DSM approach \cite{germanoetal1991,lilly1992}; namely, we calculate the anisotropic stress (\ref{eq:5}) by using the following eddy viscosity
\begin{gather}
\nu^\mathrm{sgs} = C \overline{\Delta}^2 \sqrt{2 \overline{s}^2}, \ \ 
C = \max \left( \frac{\langle L_{ij} M_{ij} \rangle_\mathrm{S}}{2 \langle M_{\ell m} M_{\ell m} \rangle_\mathrm{S}}, 0 \right),
\label{eq:22}
\end{gather}
instead of using Eq.~(\ref{eq:4}). Here, $L_{ij} = \widehat{\overline{u}_i \overline{u}_j} - \widehat{\overline{u}}_i \widehat{\overline{u}}_j$, $M_{ij} = \widehat{\overline{\Delta}}^2 \sqrt{ 2 \widehat{\overline{s}}^2} \ \widehat{\overline{s}}_{ij} - \overline{\Delta}^2 \widehat{ \sqrt{2 \overline{s}^2} \ \overline{s}_{ij}}$, and $\langle \cdot \rangle_\mathrm{S}$ represents the averaging over the $x$-$z$ plane. We clip the negative eddy viscosity similarly to that in \textit{a posteriori} tests of DSM. In contrast to the anisotropic stress based on Eq.~(\ref{eq:4}), that based on Eq.~(\ref{eq:22}) allows the energy transfer between GS and SGS. When the anisotropic redistribution based on Eq.~(\ref{eq:22}) (hereafter referred to as AR-DSM) becomes negligibly small, we can infer that the DSM predicts well the interaction between the GS velocity fluctuation and SGS even in the \textit{a posteriori} test. In other words, when the AR-DSM significantly contributes to the budgets, we can infer that the anisotropic stress is necessary even when we employ a conventional eddy-viscosity model in \textit{a posteriori} tests. Furthermore, when the contribution of AR-DSM is positive in the budgets, it cannot be modeled in terms of the eddy-viscosity assumption.

\begin{figure}[tb]
\centering
\includegraphics[width=\textwidth]{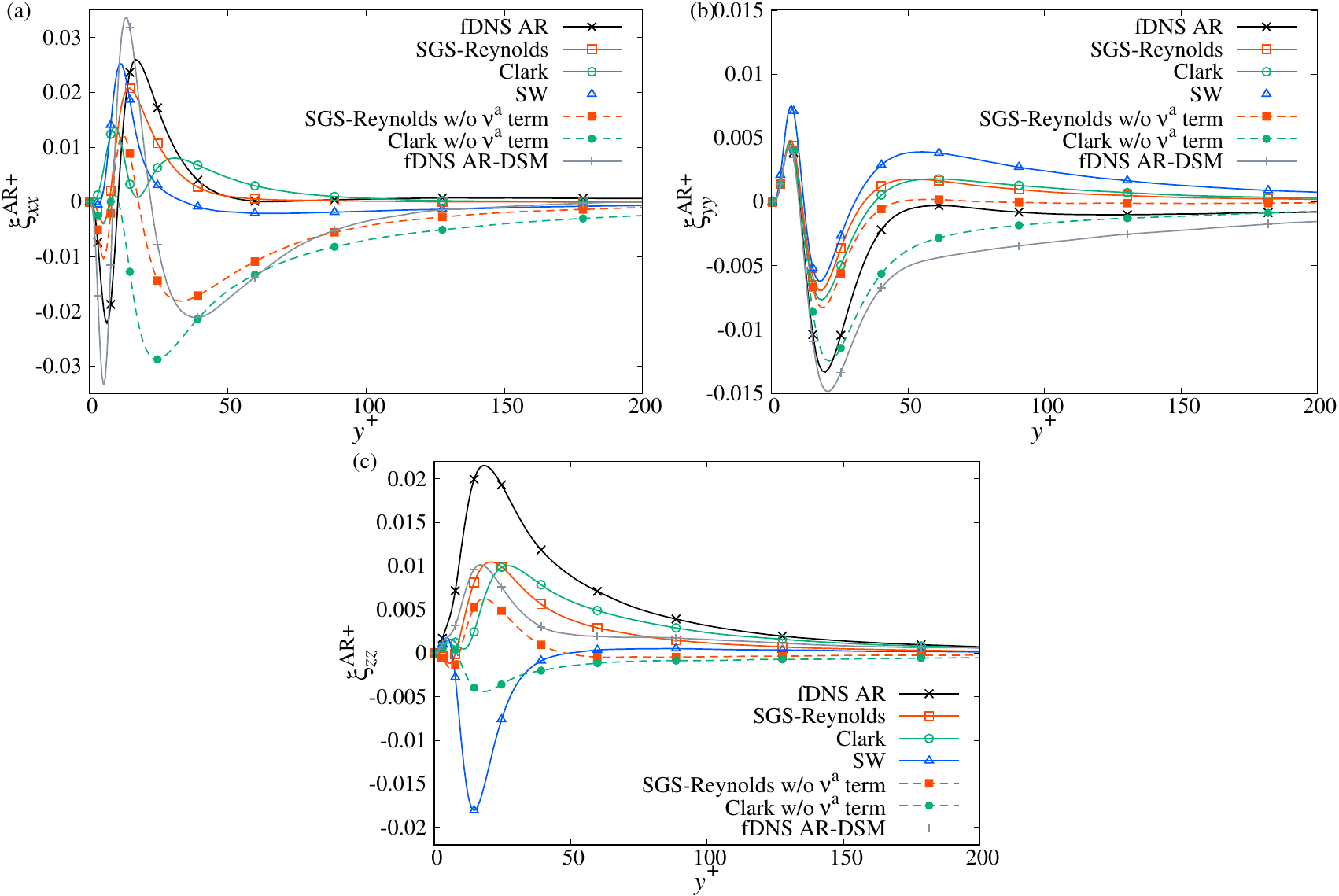}
\caption{\textit{A priori} test of the anisotropic redistribution term for (a) streamwise (b) wall-normal, and (c) spanwise components for the coarse filter case. We also plot the models without the $\nu^\mathrm{a}$-related term of Eqs.~(\ref{eq:19}) and (\ref{eq:20}) in dashed lines.}
\label{fig:13}
\end{figure}

Figure~\ref{fig:13} shows the \textit{a priori} prediction of the normal components of the anisotropic redistribution. We refer to the models provided in Eqs.~(\ref{eq:19}), (\ref{eq:20}), and (\ref{eq:21}) as SGS-Reynolds, Clark, and SW, respectively. For reference, we also plot models (\ref{eq:19}) and (\ref{eq:20}) without the $\nu^\mathrm{a}$-related term (namely, $\tau^\mathrm{ani}_{ij} = \tau^\mathrm{sgs}_{\ell \ell} \tau^\mathrm{a}_{ij}|_\mathrm{tl}/\tau^\mathrm{a}_{mm}$) where they are allowed to contribute to the energy transfer including backward scatter. For the streamwise component, all three models predict a positive contribution, although they fail to predict a negative contribution observed in fDNS in the vicinity of the wall. The SGS-Reynolds model provides a profile most similar to the fDNS, whereas the Clark model shows a different profile. The models without the $\nu^\mathrm{a}$ term provide a strongly negative profile that is far from that of the fDNS. 
The profile of AR-DSM is rather similar to that of EV dissipation in Fig.~\ref{fig:6}(b). Namely, it is almost negative except near a positive peak at $y^+ \approx 10$. The value of AR-DSM at $y^+ \approx 40$ is about $-0.02$, which is the half of EV dissipation of fDNS there (see Fig.~\ref{fig:6}(b)). Note that the anisotropic redistribution of fDNS is negligibly small at the point, and thus the total dissipation rate of $R^\mathrm{GS}_{xx}$ due to the SGS stress is twice that estimated by DSM in the \textit{a priori} test. In other words, the eddy viscosity based on the DSM is less dissipative than the ideal one given by Eq.~(\ref{eq:4}).

The wall-normal component shown in Fig.~\ref{fig:13}(b) is well predicted by all models including those without the $\nu^\mathrm{a}$ term. 
The profile of AR-DSM also agrees with that of fDNS based on Eq.~(\ref{eq:4}) qualitatively.
For the spanwise component shown in Fig.~\ref{fig:13}(c), the SW model provides a different sign than the fDNS does, which indicates that it attenuates the spanwise GS velocity fluctuation. The SGS-Reynolds and Clark models succeed in predicting a positive profile observed in the fDNS, although their intensity is small. In contrast to the Clark model, the SGS-Reynolds model can provide a positive contribution without the $\nu^\mathrm{a}$ term. 
The positive value of AR-DSM is smaller than that of fDNS based on Eq.~(\ref{eq:4}) because the eddy viscosity based on the DSM is less dissipative than that given by Eq.~(\ref{eq:4}) as mentioned above. If we employ a sufficiently dissipative eddy-viscosity model, the positive contribution of anisotropic redistribution becomes large because the sum of EV dissipation and anisotropic redistribution terms is positive at $y^+ \approx 20$ for the fDNS as shown in Fig.~\ref{fig:8}. Therefore, we can infer that regardless of the choice of the eddy-viscosity model, the anisotropic stress has a significant contribution to the budgets even when we employ a conventional eddy-viscosity model in \textit{a posteriori} tests of coarse grid LES.

In general, the models with the $\nu^\mathrm{a}$ term provide better results than those without it. Although the $\nu^\mathrm{a}$ term is an artificial term introduced to remove energy transfer, it can improve the prediction of the statistics. However, we do not consider that the models with $\nu^\mathrm{a}$ term are always superior to other models in predicting the budget. Furthermore, in the \textit{a posteriori} test, we must determine the closed expression of $\nu^\mathrm{sgs}$ without using DNS data. The \textit{a posteriori} performance of the model depends on the combination of the models of $\nu^\mathrm{sgs}$ and $\tau^\mathrm{ani}_{ij}$. The main conclusion of this study is that the anisotropic stress that predicts a positive contribution to the spanwise GS velocity fluctuation is key to improving SGS models.

\section{\label{sec:level5}Conclusions}

We have investigated the budget equation for grid-scale (GS or resolved scale) Reynolds stress in turbulent channel flows. In the analysis, we have decomposed the subgrid-scale (SGS) stress into two parts: the isotropic eddy-viscosity term, which governs energy transfer between GS and SGS, and the anisotropic term, which is separated from the energy transfer. According to this decomposition, the SGS dissipation is decomposed into eddy-viscosity dissipation and anisotropic redistribution terms. To clearly observe the role of the anisotropy of SGS stress, we have employed a coarse-size filter in addition to a medium-size filter. The filter length is chosen such that the conventional eddy-viscosity models can fairly predict the mean velocity profile for the medium filter case, whereas it fails for the coarse filter case (see Appendix~\ref{sec:b}).

The contribution of anisotropic redistribution to the budget of GS turbulent kinetic energy for the medium filter is negligible, whereas in the coarse filter, it has a small but positive contribution. A similar effect has been observed for the streamwise component of GS Reynolds stress. For the wall-normal component of GS Reynolds stress budget, the anisotropic redistribution is negative in the entire region for both filter sizes. In contrast, for the spanwise component of GS Reynolds stress budget, the anisotropic redistribution is always positive for both filter sizes. Furthermore, the SGS dissipation, which is the sum of eddy-viscosity dissipation and anisotropic redistribution, is also positive in the near-wall region. For the coarse filter case, the contribution of anisotropic redistribution is comparable to that of pressure redistribution. Therefore, anisotropic SGS stress is indispensable for reproducing the generation of GS spanwise velocity fluctuation for LES using a coarse grid resolution.

It has been suggested that the positive contribution of SGS dissipation to the budget is related to the coherent structures in the near-wall region of turbulent shear flows \cite{harteletal1994,piomellietal1996,cda2012,hamba2019}. To determine the relationship between coherent structures and positive contribution of anisotropic redistribution to the budget, we have performed a spectral analysis of the GS Reynolds stress budget. For the coarse filter case, the anisotropic redistribution is positive in a relatively small-scale region in the near-wall region of the budget for the streamwise component of GS Reynolds stress spectrum. This trend is consistent with the streak breakdown process in the self-sustaining process of wall-bounded turbulent shear flows \cite{hamiltonetal1995,waleffe1997}. In the budget for spanwise GS Reynolds stress spectrum in the near-wall region, the anisotropic redistribution has a positive contribution where the streamwise length scale is large and the spanwise length scale is close to the typical spacing of streaks. The spanwise velocity fluctuation accompanied by a nonzero spanwise wavenumber is related to streamwise vortices. Therefore, we can infer that the positive contribution of anisotropic redistribution to the spanwise GS Reynolds stress is related to the generation of streamwise vortices in the self-sustaining process \cite{hamiltonetal1995,waleffe1997}. Thus, this study suggests that the anisotropic part of SGS stress is responsible for the generation of coherent structures in wall-bounded turbulent shear flows.

We have performed an \textit{a priori} test of several existing models of anisotropic stress in terms of anisotropic redistribution. Among them, the model expression employed in the stabilized mixed model \cite{abe2013} seems to be the best. However, the intensity of the spanwise component is small. In addition, the artificially introduced part that removes the energy transfer due to the scale-similarity model contributes to the improvement of the profile of anisotropic redistribution. A quadratic nonlinear model based on the velocity gradient \cite{marstorpetal2009,montecchiaetal2017} cannot reproduce the positive contribution to spanwise GS Reynolds stress. In the future, we intend to develop a model that predicts all the components of anisotropic redistribution. In conclusion, this study has suggested that anisotropic SGS stress reproducing a positive contribution to the GS Reynolds stress budget is key to improving SGS models. This viewpoint can be a novel guiding principle in SGS modeling, particularly for coarse grid cases.

In this study, we have not considered the filter in the wall-normal direction. When we apply the filter in the wall-normal direction, the velocity gradient in the vicinity of the wall is smoothed. Therefore, the interaction between the mean and turbulent fields should be changed. To discuss the physics of coarse LES based on the DNS in more detail, we have to consider wall-normal filtering.
However, to realize the filter that coincides with the actual LES, we have to apply the inhomogeneous filter where the filter length changes against the distance from the wall. This inhomogeneous filter is not commutative with a differential operation and induces several additional terms which are different from the stress term in the filtered continuity and Navier-Stokes equations \cite{moseretal2021}. We have to assess the effect of such terms arising from the commutation error. Furthermore, in wall-bounded turbulent flows, the interaction between near-wall small eddies and the bulk flow is also important \cite{cimarellietal2016}. The physics of wall-normal filtered velocity fields and its subgrid-scale modeling should be discussed in a future study.

\begin{acknowledgments}
K.I. was supported by a Grant-in-Aid for JSPS Fellows Grant No. JP21J00580. H.K. was supported by Keio Gijuku Academic Development Funds.
\end{acknowledgments}

\appendix

\makeatletter
\renewcommand{\theequation}{\thesection\arabic{equation}}
\@addtoreset{equation}{section}
\makeatother

\section{\label{sec:a}Decomposition of SGS diffusion term}

Similar to the SGS dissipation term, we can decompose the SGS diffusion term (\ref{eq:7h}) as follows:
\begin{align}
D^\mathrm{SGS}_{ij} & = D^\mathrm{EV}_{ij} + D^\mathrm{AS}_{ij}, 
\label{eq:b1}
\end{align}
where
\begin{subequations}
\begin{align}
D^\mathrm{EV}_{ij} & = 2 \frac{\partial}{\partial x_\ell} \left< \nu^\mathrm{sgs} \left( \overline{s}_{i\ell} \overline{u}_j' + \overline{s}_{j\ell} \overline{u}_i' \right) \right>,
\label{eq:b2a} \\
D^\mathrm{AS}_{ij} & = - \frac{\partial}{\partial x_\ell} \left< \tau^\mathrm{ani}_{i\ell} \overline{u}_j' + \tau^\mathrm{ani}_{j\ell} \overline{u}_i' \right>.
\label{eq:b2b}
\end{align}
\end{subequations}
These two terms are referred to as eddy-viscosity and anisotropic stress diffusions, respectively.

\begin{figure}[tb]
\centering
\includegraphics[width=\textwidth]{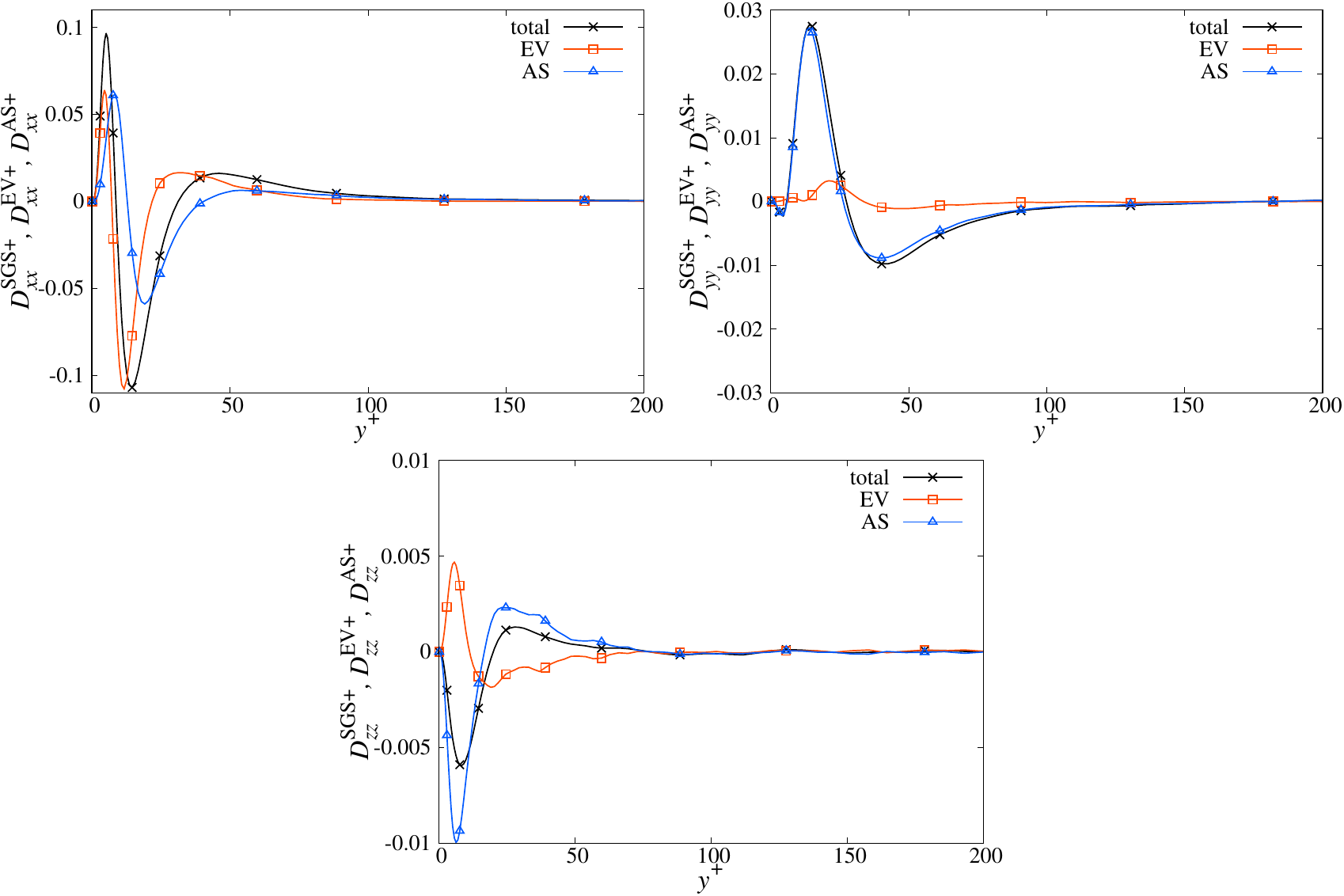}
\caption{Decomposition of SGS diffusion term for (a) streamwise (b) wall-normal, and (c) spanwise components for the coarse filter case.}
\label{fig:14}
\end{figure}

Figure~\ref{fig:14} shows the profiles of decomposed SGS diffusion terms. For the streamwise component, the eddy-viscosity diffusion (EV) provides a profile similar to SGS diffusion itself. However, the anisotropic stress diffusion (AS) also has a value comparable to that of the eddy-viscosity term and slightly alters the profile of SGS diffusion. The wall-normal component of SGS diffusion is determined by anisotropic stress diffusion, whereas the eddy-viscosity diffusion is negligible. For the spanwise component, the anisotropic stress diffusion provides a profile similar to that of SGS diffusion. The eddy-viscosity diffusion has the opposite sign to that of SGS and anisotropic stress diffusions. However, for the spanwise component, the contribution of SGS diffusion to the budget is negligible when compared with the other terms as shown in Fig.~\ref{fig:8}. In contrast, the SGS diffusion contributes significantly to the budget in the near-wall region for the wall-normal component as shown in Fig.~\ref{fig:7}. Therefore, anisotropic SGS stress is also important to predict SGS diffusion, particularly for the wall-normal component.

\section{\label{sec:b}Validation of DNS}

\begin{figure}[tb]
\centering
\includegraphics[width=\textwidth]{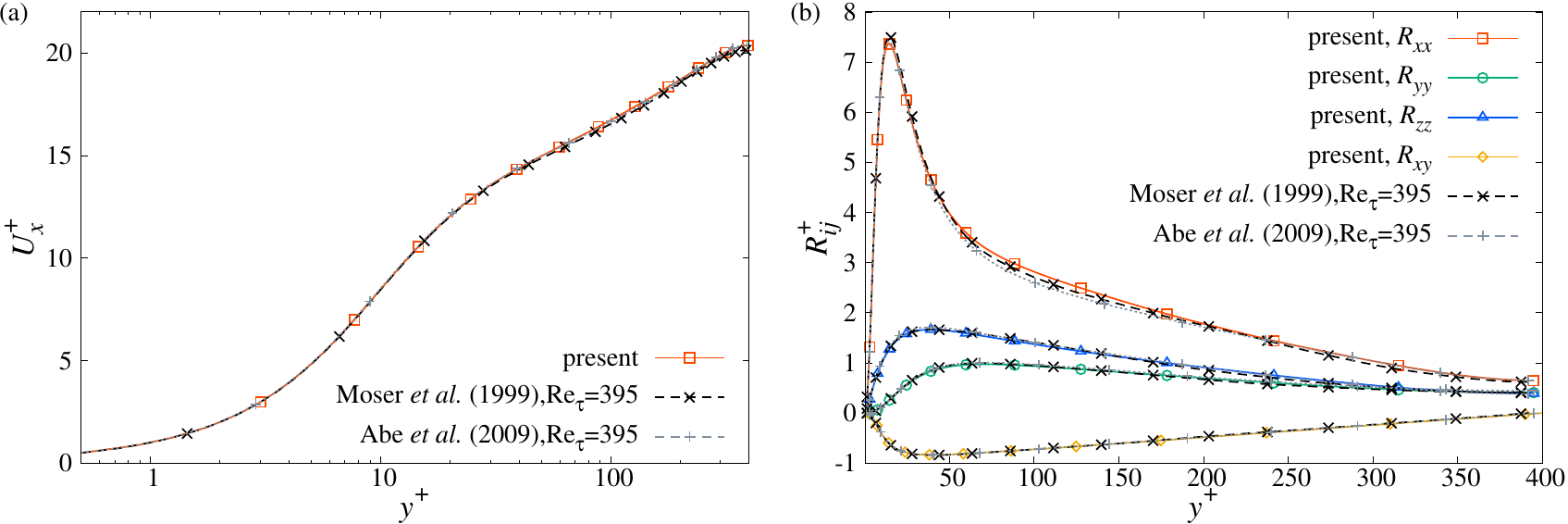}
\caption{Comparison between the present DNS and exiting database: profiles of (a) mean velocity and (b) Reynolds stress.}
\label{fig:15}
\end{figure}

Figure~\ref{fig:15} shows the comparison between the present DNS and existing ones performed by Moser \textit{et al.} \cite{mkm1999} and Abe \textit{et al.} \cite{abeetal2009}. The database of the nearest Reynolds number is chosen. Note that Abe \textit{et al.} \cite{abeetal2009} used the fourth-order finite difference in the $x$ and $z$ directions and the second-order finite difference in the $y$ direction, which are the same condition as our code, whereas Moser \textit{et al.} \cite{mkm1999} used the spectral discretization. The mean velocity profiles are almost consistent for all codes. For the Reynolds stress, the present DNS slightly overestimates the streamwise component in $50 < y^+ < 250$. However, the near-wall region seems to be well resolved. Therefore, we infer that we can analyze the physics of this flow by our present simulation, although the resolution is slightly coarser than that of references.

\section{\label{sec:c}LES results of reference filter sizes}

To observe the resolution dependence of eddy-viscosity models in the \textit{a posteriori} test, we performed LES using the dynamic Smagorinsky model (DSM) \cite{germanoetal1991,lilly1992}. The numerical method and Reynolds number are the same as the DNS provided in Sec.~\ref{sec:level3a}. We chose a grid resolution corresponding to the selected filter lengths. Namely, we performed the medium $(\Delta x^+, \Delta z^+) = (52.4, 26.2)$ and coarse $(\Delta x^+, \Delta z^+) = (105, 52.4)$ cases. We additionally performed a fine case with $(\Delta x^+, \Delta z^+) = (26.2, 13.1)$ to observe the detailed dependence on resolution. The grid number in the $y$ direction is fixed at $N_y = 96$ for all the cases. The near-wall region is well resolved as $\Delta y^+ < 1$ for the first grid from the wall, although the resolution is slightly coarser than that of DNS. The test filter operation used to evaluate the eddy viscosity in the dynamic model is calculated by retaining the first-order in the Taylor expansion (see Ref.~\cite{ik2020}) and is applied in all three directions. The filter length for the test filter is set to twice the grid width in each direction.

\begin{table}[t]
\centering
\caption{Numerical parameters for LES of DSM.}
\begin{ruledtabular}
\begin{tabular}{lcccc}
Case & $N_x \times N_y \times N_z$ & $\Delta x^+$ & $\Delta y^+$ & $\Delta z^+$ \\ \hline
fine & $96 \times 96 \times 96$ & 26.2 & 0.7--20 & 13.1 \\
medium & $48 \times 96 \times 48$ & 52.4 & 0.7--20 & 26.2 \\
coarse & $24 \times 96 \times 24$ & 105 & 0.7--20 & 52.4 \\
\end{tabular}
\end{ruledtabular}
\label{tb:1}
\end{table}

\begin{figure}[tb]
\centering
\includegraphics[width=0.6\textwidth]{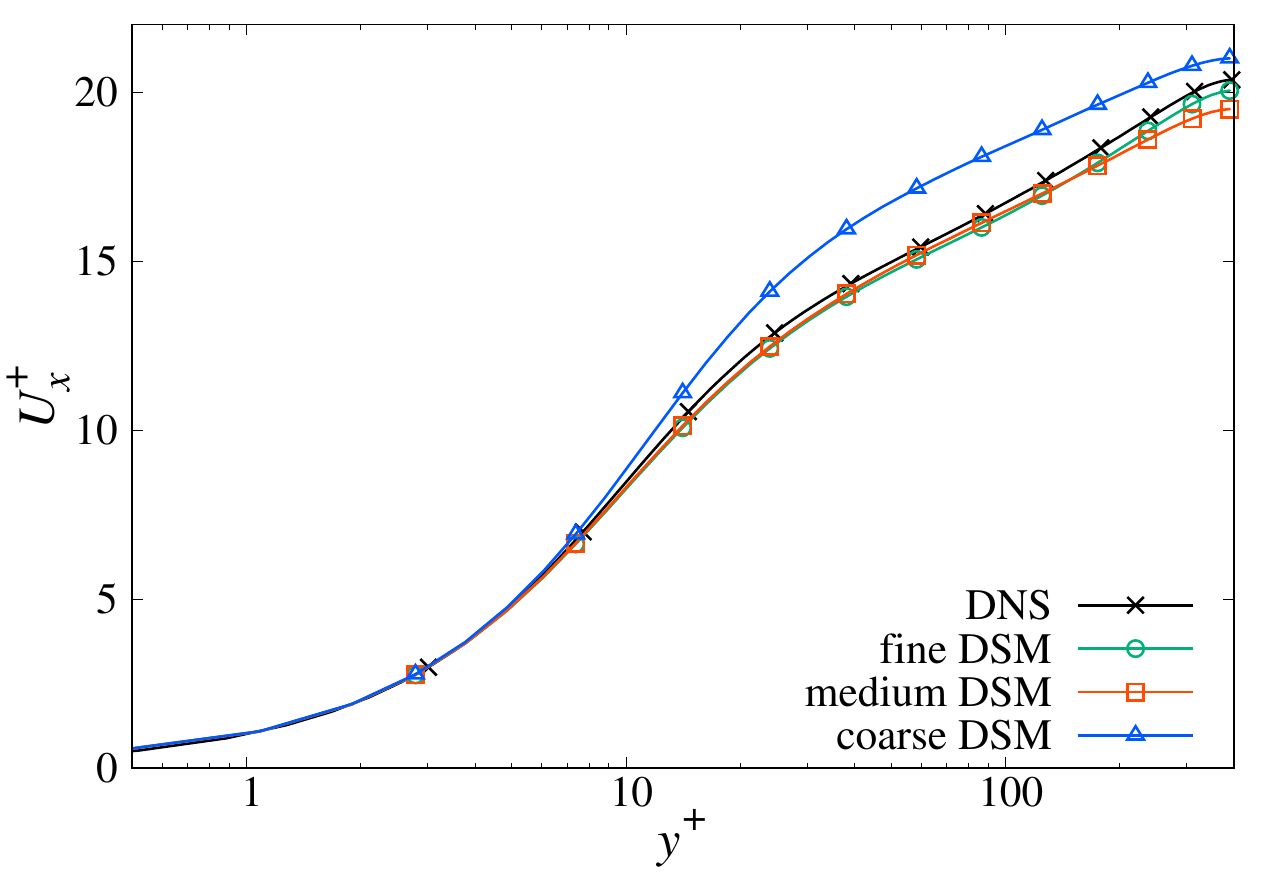}
\caption{Mean velocity profile for the LES of DSM with various grid resolutions in comparison with that for DNS.}
\label{fig:16}
\end{figure}

Figure~\ref{fig:16} shows the mean velocity profile for the LES of DSM in comparison with that for DNS. Both the fine and medium grid cases predict the mean velocity of the DNS efficiently, whereas the coarse grid case overestimates it. Thus, we can conclude that the medium grid resolution is sufficient for predicting the mean velocity profile in turbulent channel flows.

\begin{figure}[t]
\centering
\includegraphics[width=\textwidth]{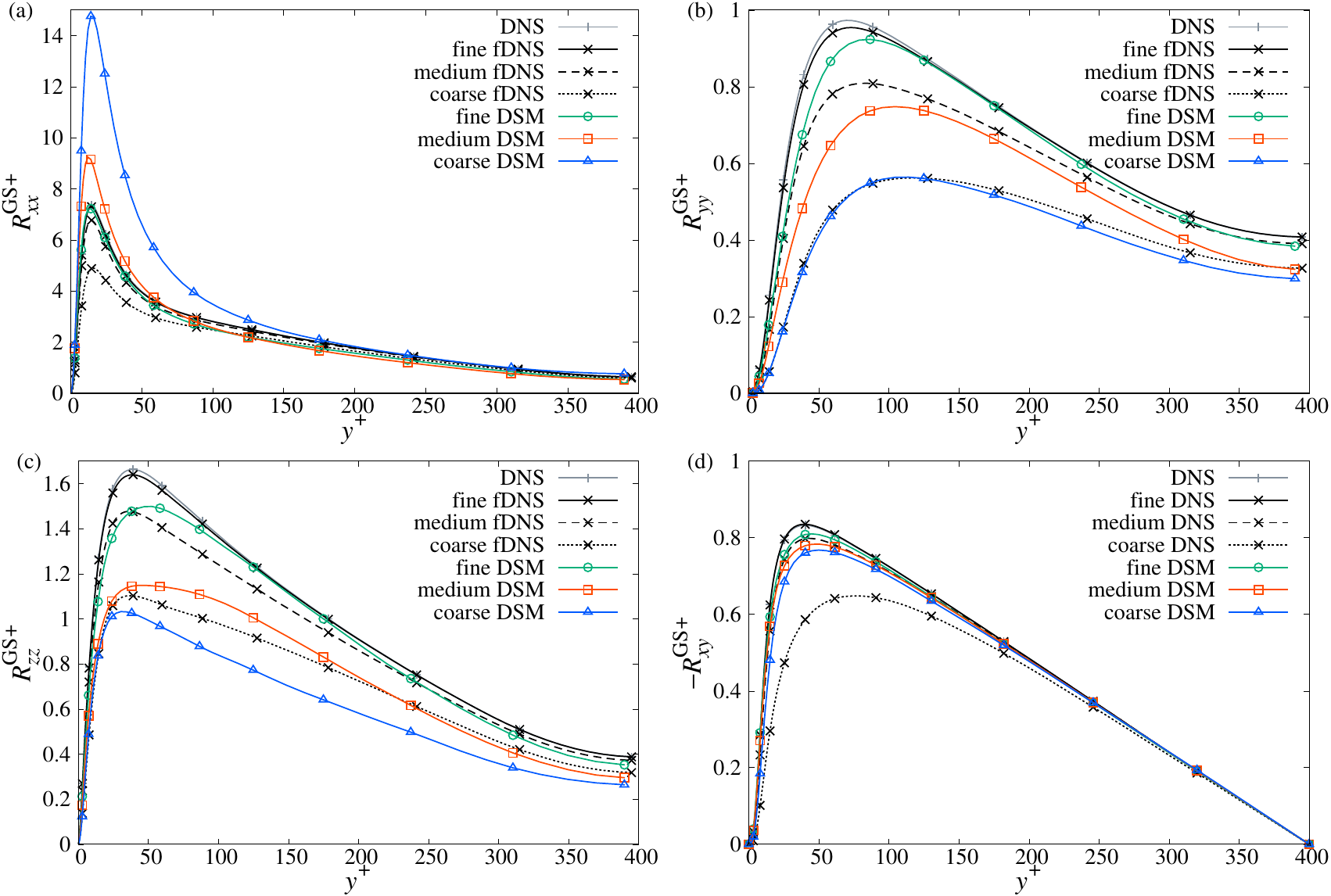}
\caption{Profiles of GS Reynolds stress for the LES of DSM in comparison with those for DNS and filtered DNS (fDNS) for (a) streamwise, (b) wall-normal, (c) spanwise, and (d) shear components.}
\label{fig:17}
\end{figure}

Figure~\ref{fig:17} shows the profiles of GS Reynolds stress for the LES of DSM in comparison with those for DNS and filtered DNS (fDNS). The fine grid case predicts all the nonzero components of GS Reynolds stress efficiently, although the wall-normal and spanwise components are slightly underestimated. In contrast, the medium grid case overestimates the streamwise component and underestimates the wall-normal and spanwise components. These trends are often observed in the LES of turbulent channel flows and are emphasized in the coarse grid case. The analysis described in Secs~\ref{sec:level3} and \ref{sec:level4} suggests that this issue can be resolved by employing an appropriate anisotropic SGS stress. The medium grid case also overestimates the shear component. However, the mean velocity is comparable to that of the DNS owing to the relatively small SGS stress (the figure is not shown).

\bibliography{ref}

\providecommand{\noopsort}[1]{}\providecommand{\singleletter}[1]{#1}%
\begin{thebibliography}{50}%
\makeatletter
\providecommand \@ifxundefined [1]{%
 \@ifx{#1\undefined}
}%
\providecommand \@ifnum [1]{%
 \ifnum #1\expandafter \@firstoftwo
 \else \expandafter \@secondoftwo
 \fi
}%
\providecommand \@ifx [1]{%
 \ifx #1\expandafter \@firstoftwo
 \else \expandafter \@secondoftwo
 \fi
}%
\providecommand \natexlab [1]{#1}%
\providecommand \enquote  [1]{``#1''}%
\providecommand \bibnamefont  [1]{#1}%
\providecommand \bibfnamefont [1]{#1}%
\providecommand \citenamefont [1]{#1}%
\providecommand \href@noop [0]{\@secondoftwo}%
\providecommand \href [0]{\begingroup \@sanitize@url \@href}%
\providecommand \@href[1]{\@@startlink{#1}\@@href}%
\providecommand \@@href[1]{\endgroup#1\@@endlink}%
\providecommand \@sanitize@url [0]{\catcode `\\12\catcode `\$12\catcode
  `\&12\catcode `\#12\catcode `\^12\catcode `\_12\catcode `\%12\relax}%
\providecommand \@@startlink[1]{}%
\providecommand \@@endlink[0]{}%
\providecommand \url  [0]{\begingroup\@sanitize@url \@url }%
\providecommand \@url [1]{\endgroup\@href {#1}{\urlprefix }}%
\providecommand \urlprefix  [0]{URL }%
\providecommand \Eprint [0]{\href }%
\providecommand \doibase [0]{https://doi.org/}%
\providecommand \selectlanguage [0]{\@gobble}%
\providecommand \bibinfo  [0]{\@secondoftwo}%
\providecommand \bibfield  [0]{\@secondoftwo}%
\providecommand \translation [1]{[#1]}%
\providecommand \BibitemOpen [0]{}%
\providecommand \bibitemStop [0]{}%
\providecommand \bibitemNoStop [0]{.\EOS\space}%
\providecommand \EOS [0]{\spacefactor3000\relax}%
\providecommand \BibitemShut  [1]{\csname bibitem#1\endcsname}%
\let\auto@bib@innerbib\@empty
\bibitem [{\citenamefont {Meneveau}(1994)}]{meneveau1994}%
  \BibitemOpen
  \bibfield  {author} {\bibinfo {author} {\bibfnamefont {C.}~\bibnamefont
  {Meneveau}},\ }\bibfield  {title} {\bibinfo {title} {Statistics of turbulence
  subgrid-scale stresses: {Necessary} conditions and experimental tests},\
  }\href@noop {} {\bibfield  {journal} {\bibinfo  {journal} {Phys. Fluids}\
  }\textbf {\bibinfo {volume} {6}},\ \bibinfo {pages} {815} (\bibinfo {year}
  {1994})}\BibitemShut {NoStop}%
\bibitem [{\citenamefont {Jim\'enez}\ and\ \citenamefont
  {Moser}(2000)}]{jm2000}%
  \BibitemOpen
  \bibfield  {author} {\bibinfo {author} {\bibfnamefont {J.}~\bibnamefont
  {Jim\'enez}}\ and\ \bibinfo {author} {\bibfnamefont {R.~D.}\ \bibnamefont
  {Moser}},\ }\bibfield  {title} {\bibinfo {title} {Large eddy simulation:
  {Where} are we and what can we expect?},\ }\href@noop {} {\bibfield
  {journal} {\bibinfo  {journal} {AIAA J.}\ }\textbf {\bibinfo {volume} {38}},\
  \bibinfo {pages} {605} (\bibinfo {year} {2000})}\BibitemShut {NoStop}%
\bibitem [{\citenamefont {Li}\ and\ \citenamefont {Meneveau}(2004)}]{lm2004}%
  \BibitemOpen
  \bibfield  {author} {\bibinfo {author} {\bibfnamefont {Y.}~\bibnamefont
  {Li}}\ and\ \bibinfo {author} {\bibfnamefont {C.}~\bibnamefont {Meneveau}},\
  }\bibfield  {title} {\bibinfo {title} {Analysis of mean momentum flux in
  subgrid models of turbulence},\ }\href@noop {} {\bibfield  {journal}
  {\bibinfo  {journal} {Phys. Fluids}\ }\textbf {\bibinfo {volume} {16}},\
  \bibinfo {pages} {3483} (\bibinfo {year} {2004})}\BibitemShut {NoStop}%
\bibitem [{\citenamefont {Marstorp}\ \emph {et~al.}(2009)\citenamefont
  {Marstorp}, \citenamefont {Brethouwer}, \citenamefont {Grundestam},\ and\
  \citenamefont {Johansson}}]{marstorpetal2009}%
  \BibitemOpen
  \bibfield  {author} {\bibinfo {author} {\bibfnamefont {L.}~\bibnamefont
  {Marstorp}}, \bibinfo {author} {\bibfnamefont {G.}~\bibnamefont
  {Brethouwer}}, \bibinfo {author} {\bibfnamefont {O.}~\bibnamefont
  {Grundestam}},\ and\ \bibinfo {author} {\bibfnamefont {A.~V.}\ \bibnamefont
  {Johansson}},\ }\bibfield  {title} {\bibinfo {title} {Explicit algebraic
  subgrid stress models with application to rotating channel flow},\
  }\href@noop {} {\bibfield  {journal} {\bibinfo  {journal} {J. Fluid Mech.}\
  }\textbf {\bibinfo {volume} {639}},\ \bibinfo {pages} {403} (\bibinfo {year}
  {2009})}\BibitemShut {NoStop}%
\bibitem [{\citenamefont {Montecchia}\ \emph {et~al.}(2017)\citenamefont
  {Montecchia}, \citenamefont {Brethouwer}, \citenamefont {Johansson},\ and\
  \citenamefont {Wallin}}]{montecchiaetal2017}%
  \BibitemOpen
  \bibfield  {author} {\bibinfo {author} {\bibfnamefont {M.}~\bibnamefont
  {Montecchia}}, \bibinfo {author} {\bibfnamefont {G.}~\bibnamefont
  {Brethouwer}}, \bibinfo {author} {\bibfnamefont {A.~V.}\ \bibnamefont
  {Johansson}},\ and\ \bibinfo {author} {\bibfnamefont {S.}~\bibnamefont
  {Wallin}},\ }\bibfield  {title} {\bibinfo {title} {Taking large-eddy
  simulation of wall-bounded flows to higher {Reynolds} numbers by use of
  anisotropy-resolving subgrid models},\ }\href@noop {} {\bibfield  {journal}
  {\bibinfo  {journal} {Phys. Rev. Fluids}\ }\textbf {\bibinfo {volume} {2}},\
  \bibinfo {pages} {034601} (\bibinfo {year} {2017})}\BibitemShut {NoStop}%
\bibitem [{\citenamefont {Abe}(2013)}]{abe2013}%
  \BibitemOpen
  \bibfield  {author} {\bibinfo {author} {\bibfnamefont {K.}~\bibnamefont
  {Abe}},\ }\bibfield  {title} {\bibinfo {title} {An improved
  anisotropy-resolving subgrid-scale model with the aid of a scale-similarity
  modeling concept},\ }\href@noop {} {\bibfield  {journal} {\bibinfo  {journal}
  {Int. J. Heat Fluid Flow}\ }\textbf {\bibinfo {volume} {39}},\ \bibinfo
  {pages} {42} (\bibinfo {year} {2013})}\BibitemShut {NoStop}%
\bibitem [{\citenamefont {Inagaki}\ and\ \citenamefont {Abe}(2017)}]{ia2017}%
  \BibitemOpen
  \bibfield  {author} {\bibinfo {author} {\bibfnamefont {M.}~\bibnamefont
  {Inagaki}}\ and\ \bibinfo {author} {\bibfnamefont {K.}~\bibnamefont {Abe}},\
  }\bibfield  {title} {\bibinfo {title} {An improved anisotropy-resolving
  subgrid-scale model for flows in laminar--turbulent transition region},\
  }\href@noop {} {\bibfield  {journal} {\bibinfo  {journal} {Int. J. Heat Fluid
  Flow}\ }\textbf {\bibinfo {volume} {64}},\ \bibinfo {pages} {137} (\bibinfo
  {year} {2017})}\BibitemShut {NoStop}%
\bibitem [{\citenamefont {Inagaki}\ and\ \citenamefont
  {Kobayashi}(2020)}]{ik2020}%
  \BibitemOpen
  \bibfield  {author} {\bibinfo {author} {\bibfnamefont {K.}~\bibnamefont
  {Inagaki}}\ and\ \bibinfo {author} {\bibfnamefont {H.}~\bibnamefont
  {Kobayashi}},\ }\bibfield  {title} {\bibinfo {title} {Role of various
  scale-similarity models in stabilized mixed subgrid-scale model},\
  }\href@noop {} {\bibfield  {journal} {\bibinfo  {journal} {Phys. Fluids}\
  }\textbf {\bibinfo {volume} {32}},\ \bibinfo {pages} {075108} (\bibinfo
  {year} {2020})}\BibitemShut {NoStop}%
\bibitem [{\citenamefont {Agrawal}\ \emph {et~al.}(2022)\citenamefont
  {Agrawal}, \citenamefont {Whitmore}, \citenamefont {Griffin}, \citenamefont
  {Bose},\ and\ \citenamefont {Moin}}]{agrawaletal2022}%
  \BibitemOpen
  \bibfield  {author} {\bibinfo {author} {\bibfnamefont {R.}~\bibnamefont
  {Agrawal}}, \bibinfo {author} {\bibfnamefont {M.~P.}\ \bibnamefont
  {Whitmore}}, \bibinfo {author} {\bibfnamefont {K.~P.}\ \bibnamefont
  {Griffin}}, \bibinfo {author} {\bibfnamefont {S.~T.}\ \bibnamefont {Bose}},\
  and\ \bibinfo {author} {\bibfnamefont {P.}~\bibnamefont {Moin}},\ }\bibfield
  {title} {\bibinfo {title} {Non-{Boussinesq} subgrid-scale model with dynamic
  tensorial coefficients},\ }\href@noop {} {\bibfield  {journal} {\bibinfo
  {journal} {Phys. Rev. Fluids}\ }\textbf {\bibinfo {volume} {7}},\ \bibinfo
  {pages} {074602} (\bibinfo {year} {2022})}\BibitemShut {NoStop}%
\bibitem [{\citenamefont {Cimarelli}\ and\ \citenamefont {{De
  Angelis}}(2014)}]{cda2014}%
  \BibitemOpen
  \bibfield  {author} {\bibinfo {author} {\bibfnamefont {A.}~\bibnamefont
  {Cimarelli}}\ and\ \bibinfo {author} {\bibfnamefont {E.}~\bibnamefont {{De
  Angelis}}},\ }\bibfield  {title} {\bibinfo {title} {The physics of energy
  transfer toward improved subgrid-scale models},\ }\href@noop {} {\bibfield
  {journal} {\bibinfo  {journal} {Phys. Fluids}\ }\textbf {\bibinfo {volume}
  {26}},\ \bibinfo {pages} {055103} (\bibinfo {year} {2014})}\BibitemShut
  {NoStop}%
\bibitem [{\citenamefont {Cimarelli}\ \emph {et~al.}(2019)\citenamefont
  {Cimarelli}, \citenamefont {Abb\`a},\ and\ \citenamefont
  {Germano}}]{cimarellietal2019}%
  \BibitemOpen
  \bibfield  {author} {\bibinfo {author} {\bibfnamefont {A.}~\bibnamefont
  {Cimarelli}}, \bibinfo {author} {\bibfnamefont {A.}~\bibnamefont {Abb\`a}},\
  and\ \bibinfo {author} {\bibfnamefont {M.}~\bibnamefont {Germano}},\
  }\bibfield  {title} {\bibinfo {title} {General formalism for a reduced
  description and modelling of momentum and energy transfer in turbulence},\
  }\href@noop {} {\bibfield  {journal} {\bibinfo  {journal} {J. Fluid Mech.}\
  }\textbf {\bibinfo {volume} {866}},\ \bibinfo {pages} {865} (\bibinfo {year}
  {2019})}\BibitemShut {NoStop}%
\bibitem [{\citenamefont {Honnert}\ \emph {et~al.}(2020)\citenamefont
  {Honnert}, \citenamefont {Efstathiou}, \citenamefont {Beare}, \citenamefont
  {Ito}, \citenamefont {Lock}, \citenamefont {Neggers}, \citenamefont {Plant},
  \citenamefont {Shin}, \citenamefont {Tomassini},\ and\ \citenamefont
  {Zhou}}]{honnertetal2020}%
  \BibitemOpen
  \bibfield  {author} {\bibinfo {author} {\bibfnamefont {R.}~\bibnamefont
  {Honnert}}, \bibinfo {author} {\bibfnamefont {G.~A.}\ \bibnamefont
  {Efstathiou}}, \bibinfo {author} {\bibfnamefont {R.~J.}\ \bibnamefont
  {Beare}}, \bibinfo {author} {\bibfnamefont {J.}~\bibnamefont {Ito}}, \bibinfo
  {author} {\bibfnamefont {A.}~\bibnamefont {Lock}}, \bibinfo {author}
  {\bibfnamefont {R.}~\bibnamefont {Neggers}}, \bibinfo {author} {\bibfnamefont
  {R.~S.}\ \bibnamefont {Plant}}, \bibinfo {author} {\bibfnamefont {H.~H.}\
  \bibnamefont {Shin}}, \bibinfo {author} {\bibfnamefont {L.}~\bibnamefont
  {Tomassini}},\ and\ \bibinfo {author} {\bibfnamefont {B.}~\bibnamefont
  {Zhou}},\ }\bibfield  {title} {\bibinfo {title} {The atmospheric boundary
  layer and the ``gray zone'' of turbulence: {A} critical review},\ }\href@noop
  {} {\bibfield  {journal} {\bibinfo  {journal} {J. Geophys. Res.}\ }\textbf
  {\bibinfo {volume} {125}},\ \bibinfo {pages} {e2019JD030317} (\bibinfo {year}
  {2020})}\BibitemShut {NoStop}%
\bibitem [{\citenamefont {Abe}(2019)}]{abe2019}%
  \BibitemOpen
  \bibfield  {author} {\bibinfo {author} {\bibfnamefont {K.}~\bibnamefont
  {Abe}},\ }\bibfield  {title} {\bibinfo {title} {Notable effect of the
  subgrid-scale stress anisotropy on mean-velocity prediction through budget of
  the grid-scale {Reynolds}-shear stress},\ }\href@noop {} {\bibfield
  {journal} {\bibinfo  {journal} {Phys. Fluids}\ }\textbf {\bibinfo {volume}
  {31}},\ \bibinfo {pages} {105103} (\bibinfo {year} {2019})}\BibitemShut
  {NoStop}%
\bibitem [{\citenamefont {Moser}\ \emph {et~al.}(2021)\citenamefont {Moser},
  \citenamefont {Haering},\ and\ \citenamefont {Yalla}}]{moseretal2021}%
  \BibitemOpen
  \bibfield  {author} {\bibinfo {author} {\bibfnamefont {R.~D.}\ \bibnamefont
  {Moser}}, \bibinfo {author} {\bibfnamefont {S.~W.}\ \bibnamefont {Haering}},\
  and\ \bibinfo {author} {\bibfnamefont {G.~R.}\ \bibnamefont {Yalla}},\
  }\bibfield  {title} {\bibinfo {title} {Statistical properties of
  subgrid-scale turbulence models},\ }\href@noop {} {\bibfield  {journal}
  {\bibinfo  {journal} {Annu. Rev. Fluid Mech.}\ }\textbf {\bibinfo {volume}
  {53}},\ \bibinfo {pages} {255} (\bibinfo {year} {2021})}\BibitemShut
  {NoStop}%
\bibitem [{\citenamefont {Haering}\ \emph {et~al.}(2019)\citenamefont
  {Haering}, \citenamefont {Lee},\ and\ \citenamefont
  {Moser}}]{haeringetal2019}%
  \BibitemOpen
  \bibfield  {author} {\bibinfo {author} {\bibfnamefont {S.~W.}\ \bibnamefont
  {Haering}}, \bibinfo {author} {\bibfnamefont {M.}~\bibnamefont {Lee}},\ and\
  \bibinfo {author} {\bibfnamefont {R.~D.}\ \bibnamefont {Moser}},\ }\bibfield
  {title} {\bibinfo {title} {Resolution-induced anisotropy in large-eddy
  simulations},\ }\href@noop {} {\bibfield  {journal} {\bibinfo  {journal}
  {Phys. Rev. Fluids}\ }\textbf {\bibinfo {volume} {4}},\ \bibinfo {pages}
  {114605} (\bibinfo {year} {2019})}\BibitemShut {NoStop}%
\bibitem [{\citenamefont {Domaradzki}\ \emph {et~al.}(1994)\citenamefont
  {Domaradzki}, \citenamefont {Liu}, \citenamefont {H{\"a}rtel},\ and\
  \citenamefont {Kleiser}}]{domaradzkietal1994}%
  \BibitemOpen
  \bibfield  {author} {\bibinfo {author} {\bibfnamefont {J.~A.}\ \bibnamefont
  {Domaradzki}}, \bibinfo {author} {\bibfnamefont {W.}~\bibnamefont {Liu}},
  \bibinfo {author} {\bibfnamefont {C.}~\bibnamefont {H{\"a}rtel}},\ and\
  \bibinfo {author} {\bibfnamefont {L.}~\bibnamefont {Kleiser}},\ }\bibfield
  {title} {\bibinfo {title} {Energy transfer in numerically simulated
  wall-bounded turbulent flows},\ }\href@noop {} {\bibfield  {journal}
  {\bibinfo  {journal} {Phys. Fluids}\ }\textbf {\bibinfo {volume} {6}},\
  \bibinfo {pages} {1583} (\bibinfo {year} {1994})}\BibitemShut {NoStop}%
\bibitem [{\citenamefont {H{\"a}rtel}\ and\ \citenamefont
  {Kleiser}(1998)}]{hk1998}%
  \BibitemOpen
  \bibfield  {author} {\bibinfo {author} {\bibfnamefont {C.}~\bibnamefont
  {H{\"a}rtel}}\ and\ \bibinfo {author} {\bibfnamefont {L.}~\bibnamefont
  {Kleiser}},\ }\bibfield  {title} {\bibinfo {title} {Analysis and modelling of
  subgrid-scale motions in near-wall turbulence},\ }\href@noop {} {\bibfield
  {journal} {\bibinfo  {journal} {J. Fluid Mech.}\ }\textbf {\bibinfo {volume}
  {356}},\ \bibinfo {pages} {327} (\bibinfo {year} {1998})}\BibitemShut
  {NoStop}%
\bibitem [{\citenamefont {Liu}\ \emph {et~al.}(1994)\citenamefont {Liu},
  \citenamefont {Meneveau},\ and\ \citenamefont {Katz}}]{liuetal1994}%
  \BibitemOpen
  \bibfield  {author} {\bibinfo {author} {\bibfnamefont {S.}~\bibnamefont
  {Liu}}, \bibinfo {author} {\bibfnamefont {C.}~\bibnamefont {Meneveau}},\ and\
  \bibinfo {author} {\bibfnamefont {J.}~\bibnamefont {Katz}},\ }\bibfield
  {title} {\bibinfo {title} {On the properties of similarity subgrid-scale
  models as deduced from measurements in a turbulent jet},\ }\href@noop {}
  {\bibfield  {journal} {\bibinfo  {journal} {J. Fluid Mech.}\ }\textbf
  {\bibinfo {volume} {275}},\ \bibinfo {pages} {83} (\bibinfo {year}
  {1994})}\BibitemShut {NoStop}%
\bibitem [{\citenamefont {Tao}\ \emph {et~al.}(2002)\citenamefont {Tao},
  \citenamefont {Katz},\ and\ \citenamefont {Meneveau}}]{taoetal2002}%
  \BibitemOpen
  \bibfield  {author} {\bibinfo {author} {\bibfnamefont {B.}~\bibnamefont
  {Tao}}, \bibinfo {author} {\bibfnamefont {J.}~\bibnamefont {Katz}},\ and\
  \bibinfo {author} {\bibfnamefont {C.}~\bibnamefont {Meneveau}},\ }\bibfield
  {title} {\bibinfo {title} {Statistical geometry of subgrid-scale stresses
  determined from holographic particle image velocimetry measurements},\
  }\href@noop {} {\bibfield  {journal} {\bibinfo  {journal} {J. Fluid Mech}\
  }\textbf {\bibinfo {volume} {457}},\ \bibinfo {pages} {35} (\bibinfo {year}
  {2002})}\BibitemShut {NoStop}%
\bibitem [{\citenamefont {Horiuti}(2003)}]{horiuti2003}%
  \BibitemOpen
  \bibfield  {author} {\bibinfo {author} {\bibfnamefont {K.}~\bibnamefont
  {Horiuti}},\ }\bibfield  {title} {\bibinfo {title} {Roles of non-aligned
  eigenvectors of strain-rate and subgrid-scale stress tensors in turbulence
  generation},\ }\href@noop {} {\bibfield  {journal} {\bibinfo  {journal} {J.
  Fluid Mech}\ }\textbf {\bibinfo {volume} {491}},\ \bibinfo {pages} {65}
  (\bibinfo {year} {2003})}\BibitemShut {NoStop}%
\bibitem [{\citenamefont {Piomelli}\ \emph {et~al.}(1991)\citenamefont
  {Piomelli}, \citenamefont {Cabot}, \citenamefont {Moin},\ and\ \citenamefont
  {Lee}}]{piomellietal1991}%
  \BibitemOpen
  \bibfield  {author} {\bibinfo {author} {\bibfnamefont {U.}~\bibnamefont
  {Piomelli}}, \bibinfo {author} {\bibfnamefont {W.~H.}\ \bibnamefont {Cabot}},
  \bibinfo {author} {\bibfnamefont {P.}~\bibnamefont {Moin}},\ and\ \bibinfo
  {author} {\bibfnamefont {S.}~\bibnamefont {Lee}},\ }\bibfield  {title}
  {\bibinfo {title} {Subgrid-scale backscatter in turbulent and transitional
  flows},\ }\href@noop {} {\bibfield  {journal} {\bibinfo  {journal} {Phys.
  Fluids A}\ }\textbf {\bibinfo {volume} {3}},\ \bibinfo {pages} {1766}
  (\bibinfo {year} {1991})}\BibitemShut {NoStop}%
\bibitem [{\citenamefont {Aoyama}\ \emph {et~al.}(2005)\citenamefont {Aoyama},
  \citenamefont {Ishihara}, \citenamefont {Kaneda}, \citenamefont {Yokokawa},
  \citenamefont {Itakura},\ and\ \citenamefont {Uno}}]{aoyamaetal2005}%
  \BibitemOpen
  \bibfield  {author} {\bibinfo {author} {\bibfnamefont {T.}~\bibnamefont
  {Aoyama}}, \bibinfo {author} {\bibfnamefont {T.}~\bibnamefont {Ishihara}},
  \bibinfo {author} {\bibfnamefont {Y.}~\bibnamefont {Kaneda}}, \bibinfo
  {author} {\bibfnamefont {M.}~\bibnamefont {Yokokawa}}, \bibinfo {author}
  {\bibfnamefont {K.}~\bibnamefont {Itakura}},\ and\ \bibinfo {author}
  {\bibfnamefont {A.}~\bibnamefont {Uno}},\ }\bibfield  {title} {\bibinfo
  {title} {Statistics of energy transfer in high-resolution direct numerical
  simulation of turbulence in a periodic box},\ }\href@noop {} {\bibfield
  {journal} {\bibinfo  {journal} {J. Phys. Soc. Jpn.}\ }\textbf {\bibinfo
  {volume} {74}},\ \bibinfo {pages} {3202} (\bibinfo {year}
  {2005})}\BibitemShut {NoStop}%
\bibitem [{\citenamefont {Bardina}\ \emph {et~al.}(1983)\citenamefont
  {Bardina}, \citenamefont {Ferziger},\ and\ \citenamefont
  {Reynolds}}]{bardinaetal1983}%
  \BibitemOpen
  \bibfield  {author} {\bibinfo {author} {\bibfnamefont {J.}~\bibnamefont
  {Bardina}}, \bibinfo {author} {\bibfnamefont {J.~H.}\ \bibnamefont
  {Ferziger}},\ and\ \bibinfo {author} {\bibfnamefont {W.~C.}\ \bibnamefont
  {Reynolds}},\ }\bibfield  {title} {\bibinfo {title} {Improved turbulence
  models based on large eddy simulation of homogenous, incompressible,
  turbulent flows},\ }\href@noop {} {\bibfield  {journal} {\bibinfo  {journal}
  {Report TF-19. Thermosciences Division, Dep. of Mech. Eng., Stanford
  University, Stanford, California}\ } (\bibinfo {year} {1983})}\BibitemShut
  {NoStop}%
\bibitem [{\citenamefont {Meneveau}\ and\ \citenamefont {Katz}(2000)}]{mk2000}%
  \BibitemOpen
  \bibfield  {author} {\bibinfo {author} {\bibfnamefont {C.}~\bibnamefont
  {Meneveau}}\ and\ \bibinfo {author} {\bibfnamefont {J.}~\bibnamefont
  {Katz}},\ }\bibfield  {title} {\bibinfo {title} {Scale-invariance and
  turbulence models for large-eddy simulation},\ }\href@noop {} {\bibfield
  {journal} {\bibinfo  {journal} {Annu. Rev. Fluid Mech.}\ }\textbf {\bibinfo
  {volume} {32}},\ \bibinfo {pages} {1} (\bibinfo {year} {2000})}\BibitemShut
  {NoStop}%
\bibitem [{\citenamefont {H{\"a}rtel}\ \emph {et~al.}(1994)\citenamefont
  {H{\"a}rtel}, \citenamefont {Kleiser}, \citenamefont {Unger},\ and\
  \citenamefont {Friedrich}}]{harteletal1994}%
  \BibitemOpen
  \bibfield  {author} {\bibinfo {author} {\bibfnamefont {C.}~\bibnamefont
  {H{\"a}rtel}}, \bibinfo {author} {\bibfnamefont {L.}~\bibnamefont {Kleiser}},
  \bibinfo {author} {\bibfnamefont {F.}~\bibnamefont {Unger}},\ and\ \bibinfo
  {author} {\bibfnamefont {R.}~\bibnamefont {Friedrich}},\ }\bibfield  {title}
  {\bibinfo {title} {Subgrid-scale energy transfer in the near-wall region of
  turbulent flows},\ }\href@noop {} {\bibfield  {journal} {\bibinfo  {journal}
  {Phys. Fluids}\ }\textbf {\bibinfo {volume} {6}},\ \bibinfo {pages} {3130}
  (\bibinfo {year} {1994})}\BibitemShut {NoStop}%
\bibitem [{\citenamefont {Cimarelli}\ and\ \citenamefont {{De
  Angelis}}(2012)}]{cda2012}%
  \BibitemOpen
  \bibfield  {author} {\bibinfo {author} {\bibfnamefont {A.}~\bibnamefont
  {Cimarelli}}\ and\ \bibinfo {author} {\bibfnamefont {E.}~\bibnamefont {{De
  Angelis}}},\ }\bibfield  {title} {\bibinfo {title} {Anisotropic dynamics and
  sub-grid energy transfer in wall-turbulence},\ }\href@noop {} {\bibfield
  {journal} {\bibinfo  {journal} {Phys. Fluids}\ }\textbf {\bibinfo {volume}
  {24}},\ \bibinfo {pages} {015102} (\bibinfo {year} {2012})}\BibitemShut
  {NoStop}%
\bibitem [{\citenamefont {Kravchenko}\ \emph {et~al.}(1996)\citenamefont
  {Kravchenko}, \citenamefont {Moin},\ and\ \citenamefont
  {Moser}}]{kravchenkoetal1996}%
  \BibitemOpen
  \bibfield  {author} {\bibinfo {author} {\bibfnamefont {A.~G.}\ \bibnamefont
  {Kravchenko}}, \bibinfo {author} {\bibfnamefont {P.}~\bibnamefont {Moin}},\
  and\ \bibinfo {author} {\bibfnamefont {R.}~\bibnamefont {Moser}},\ }\bibfield
   {title} {\bibinfo {title} {Zonal embedded grids for numerical simulations of
  wall-bounded turbulent flows},\ }\href@noop {} {\bibfield  {journal}
  {\bibinfo  {journal} {J. Comput. Phys.}\ }\textbf {\bibinfo {volume} {127}},\
  \bibinfo {pages} {412} (\bibinfo {year} {1996})}\BibitemShut {NoStop}%
\bibitem [{\citenamefont {Morinishi}\ and\ \citenamefont
  {Vasilyev}(2001)}]{mv2001}%
  \BibitemOpen
  \bibfield  {author} {\bibinfo {author} {\bibfnamefont {Y.}~\bibnamefont
  {Morinishi}}\ and\ \bibinfo {author} {\bibfnamefont {O.~V.}\ \bibnamefont
  {Vasilyev}},\ }\bibfield  {title} {\bibinfo {title} {A recommended
  modification to the dynamic two-parameter mixed subgrid scale model for large
  eddy simulation of wall bounded turbulent flow},\ }\href@noop {} {\bibfield
  {journal} {\bibinfo  {journal} {Phys. Fluids}\ }\textbf {\bibinfo {volume}
  {13}},\ \bibinfo {pages} {3400} (\bibinfo {year} {2001})}\BibitemShut
  {NoStop}%
\bibitem [{\citenamefont {Choi}\ and\ \citenamefont {Moin}(2012)}]{cm2012}%
  \BibitemOpen
  \bibfield  {author} {\bibinfo {author} {\bibfnamefont {H.}~\bibnamefont
  {Choi}}\ and\ \bibinfo {author} {\bibfnamefont {P.}~\bibnamefont {Moin}},\
  }\bibfield  {title} {\bibinfo {title} {Grid-point requirements for large eddy
  simulation: {Chapman}'s estimates revisited},\ }\href@noop {} {\bibfield
  {journal} {\bibinfo  {journal} {Phys. Fluids}\ }\textbf {\bibinfo {volume}
  {24}},\ \bibinfo {pages} {011702} (\bibinfo {year} {2012})}\BibitemShut
  {NoStop}%
\bibitem [{\citenamefont {Kawata}\ and\ \citenamefont
  {Alfredsson}(2018)}]{ka2018}%
  \BibitemOpen
  \bibfield  {author} {\bibinfo {author} {\bibfnamefont {T.}~\bibnamefont
  {Kawata}}\ and\ \bibinfo {author} {\bibfnamefont {P.~H.}\ \bibnamefont
  {Alfredsson}},\ }\bibfield  {title} {\bibinfo {title} {Inverse interscale
  transport of the {Reynolds} shear stress in plane {Couette} turbulence},\
  }\href@noop {} {\bibfield  {journal} {\bibinfo  {journal} {Phys. Rev. Lett.}\
  }\textbf {\bibinfo {volume} {120}},\ \bibinfo {pages} {244501} (\bibinfo
  {year} {2018})}\BibitemShut {NoStop}%
\bibitem [{\citenamefont {Lee}\ and\ \citenamefont {Moser}(2019)}]{lm2019}%
  \BibitemOpen
  \bibfield  {author} {\bibinfo {author} {\bibfnamefont {M.}~\bibnamefont
  {Lee}}\ and\ \bibinfo {author} {\bibfnamefont {R.~D.}\ \bibnamefont
  {Moser}},\ }\bibfield  {title} {\bibinfo {title} {Spectral analysis of the
  budget equation in turbulent channel flows at high {Reynolds} number},\
  }\href@noop {} {\bibfield  {journal} {\bibinfo  {journal} {J. Fluid Mech.}\
  }\textbf {\bibinfo {volume} {860}},\ \bibinfo {pages} {886} (\bibinfo {year}
  {2019})}\BibitemShut {NoStop}%
\bibitem [{\citenamefont {Kajishima}\ and\ \citenamefont
  {Taira}(2017)}]{kajishimatairabook}%
  \BibitemOpen
  \bibfield  {author} {\bibinfo {author} {\bibfnamefont {T.}~\bibnamefont
  {Kajishima}}\ and\ \bibinfo {author} {\bibfnamefont {K.}~\bibnamefont
  {Taira}},\ }\href@noop {} {\emph {\bibinfo {title} {Computational Fluid
  Dynamics}}}\ (\bibinfo  {publisher} {Springer},\ \bibinfo {address} {New
  York},\ \bibinfo {year} {2017})\BibitemShut {NoStop}%
\bibitem [{\citenamefont {Cimarelli}\ \emph {et~al.}(2016)\citenamefont
  {Cimarelli}, \citenamefont {{De Angelis}}, \citenamefont {Jim\'enez},\ and\
  \citenamefont {Casciola}}]{cimarellietal2016}%
  \BibitemOpen
  \bibfield  {author} {\bibinfo {author} {\bibfnamefont {A.}~\bibnamefont
  {Cimarelli}}, \bibinfo {author} {\bibfnamefont {E.}~\bibnamefont {{De
  Angelis}}}, \bibinfo {author} {\bibfnamefont {J.}~\bibnamefont {Jim\'enez}},\
  and\ \bibinfo {author} {\bibfnamefont {C.~M.}\ \bibnamefont {Casciola}},\
  }\bibfield  {title} {\bibinfo {title} {Cascades and wall-normal fluxes in
  turbulent channel flows},\ }\href@noop {} {\bibfield  {journal} {\bibinfo
  {journal} {J. Fluid Mech.}\ }\textbf {\bibinfo {volume} {796}},\ \bibinfo
  {pages} {417} (\bibinfo {year} {2016})}\BibitemShut {NoStop}%
\bibitem [{\citenamefont {Pope}(2000)}]{popebook}%
  \BibitemOpen
  \bibfield  {author} {\bibinfo {author} {\bibfnamefont {S.~B.}\ \bibnamefont
  {Pope}},\ }\href@noop {} {\emph {\bibinfo {title} {Turbulent Flows}}}\
  (\bibinfo  {publisher} {Cambridge University Press},\ \bibinfo {address}
  {Cambridge},\ \bibinfo {year} {2000})\BibitemShut {NoStop}%
\bibitem [{\citenamefont {Lee}\ and\ \citenamefont {Moser}(2015)}]{lm2015}%
  \BibitemOpen
  \bibfield  {author} {\bibinfo {author} {\bibfnamefont {M.}~\bibnamefont
  {Lee}}\ and\ \bibinfo {author} {\bibfnamefont {R.~D.}\ \bibnamefont
  {Moser}},\ }\bibfield  {title} {\bibinfo {title} {Direct numerical simulation
  of turbulent channel flow up to {$Re_\tau \approx 5200$}},\ }\href@noop {}
  {\bibfield  {journal} {\bibinfo  {journal} {J. Fluid Mech.}\ }\textbf
  {\bibinfo {volume} {774}},\ \bibinfo {pages} {395} (\bibinfo {year}
  {2015})}\BibitemShut {NoStop}%
\bibitem [{\citenamefont {Cimarelli}\ and\ \citenamefont {{De
  Angelis}}(2011)}]{cda2011}%
  \BibitemOpen
  \bibfield  {author} {\bibinfo {author} {\bibfnamefont {A.}~\bibnamefont
  {Cimarelli}}\ and\ \bibinfo {author} {\bibfnamefont {E.}~\bibnamefont {{De
  Angelis}}},\ }\bibfield  {title} {\bibinfo {title} {Analysis of the
  kolmogorov equation for filtered wall-turbulent flows},\ }\href@noop {}
  {\bibfield  {journal} {\bibinfo  {journal} {J. Fluid Mech.}\ }\textbf
  {\bibinfo {volume} {676}},\ \bibinfo {pages} {376} (\bibinfo {year}
  {2011})}\BibitemShut {NoStop}%
\bibitem [{\citenamefont {Piomelli}\ \emph {et~al.}(1996)\citenamefont
  {Piomelli}, \citenamefont {Yu},\ and\ \citenamefont
  {Adrian}}]{piomellietal1996}%
  \BibitemOpen
  \bibfield  {author} {\bibinfo {author} {\bibfnamefont {U.}~\bibnamefont
  {Piomelli}}, \bibinfo {author} {\bibfnamefont {Y.}~\bibnamefont {Yu}},\ and\
  \bibinfo {author} {\bibfnamefont {R.~J.}\ \bibnamefont {Adrian}},\ }\bibfield
   {title} {\bibinfo {title} {Subgrid-scale energy transfer and near-wall
  turbulence structure},\ }\href@noop {} {\bibfield  {journal} {\bibinfo
  {journal} {Phys. Fluids}\ }\textbf {\bibinfo {volume} {8}},\ \bibinfo {pages}
  {215} (\bibinfo {year} {1996})}\BibitemShut {NoStop}%
\bibitem [{\citenamefont {Hamba}(2019)}]{hamba2019}%
  \BibitemOpen
  \bibfield  {author} {\bibinfo {author} {\bibfnamefont {F.}~\bibnamefont
  {Hamba}},\ }\bibfield  {title} {\bibinfo {title} {Inverse energy cascade and
  vortical structure in the near-wall region of turbulent channel flow},\
  }\href@noop {} {\bibfield  {journal} {\bibinfo  {journal} {Phys. Rev.
  Fluids}\ }\textbf {\bibinfo {volume} {4}},\ \bibinfo {pages} {114609}
  (\bibinfo {year} {2019})}\BibitemShut {NoStop}%
\bibitem [{\citenamefont {Kline}\ \emph {et~al.}(1967)\citenamefont {Kline},
  \citenamefont {Reynolds}, \citenamefont {Schraub},\ and\ \citenamefont
  {Runstadler}}]{klineetal1967}%
  \BibitemOpen
  \bibfield  {author} {\bibinfo {author} {\bibfnamefont {S.~J.}\ \bibnamefont
  {Kline}}, \bibinfo {author} {\bibfnamefont {W.~C.}\ \bibnamefont {Reynolds}},
  \bibinfo {author} {\bibfnamefont {F.~A.}\ \bibnamefont {Schraub}},\ and\
  \bibinfo {author} {\bibfnamefont {P.~W.}\ \bibnamefont {Runstadler}},\
  }\bibfield  {title} {\bibinfo {title} {The structure of turbulent boundary
  layers},\ }\href@noop {} {\bibfield  {journal} {\bibinfo  {journal} {J. Fluid
  Mech.}\ }\textbf {\bibinfo {volume} {30}},\ \bibinfo {pages} {741} (\bibinfo
  {year} {1967})}\BibitemShut {NoStop}%
\bibitem [{\citenamefont {Jim\'enez}\ and\ \citenamefont
  {Moin}(1991)}]{jm1991}%
  \BibitemOpen
  \bibfield  {author} {\bibinfo {author} {\bibfnamefont {J.}~\bibnamefont
  {Jim\'enez}}\ and\ \bibinfo {author} {\bibfnamefont {P.}~\bibnamefont
  {Moin}},\ }\bibfield  {title} {\bibinfo {title} {The minimal flow unit in
  near-wall turbulence},\ }\href@noop {} {\bibfield  {journal} {\bibinfo
  {journal} {J. Fluid Mech.}\ }\textbf {\bibinfo {volume} {225}},\ \bibinfo
  {pages} {213} (\bibinfo {year} {1991})}\BibitemShut {NoStop}%
\bibitem [{\citenamefont {Hamilton}\ \emph {et~al.}(1995)\citenamefont
  {Hamilton}, \citenamefont {Kim},\ and\ \citenamefont
  {Waleffe}}]{hamiltonetal1995}%
  \BibitemOpen
  \bibfield  {author} {\bibinfo {author} {\bibfnamefont {J.~M.}\ \bibnamefont
  {Hamilton}}, \bibinfo {author} {\bibfnamefont {J.}~\bibnamefont {Kim}},\ and\
  \bibinfo {author} {\bibfnamefont {F.}~\bibnamefont {Waleffe}},\ }\bibfield
  {title} {\bibinfo {title} {Regeneration mechanisms of near-wall turbulence
  structures},\ }\href@noop {} {\bibfield  {journal} {\bibinfo  {journal} {J.
  Fluid Mech.}\ }\textbf {\bibinfo {volume} {287}},\ \bibinfo {pages} {317}
  (\bibinfo {year} {1995})}\BibitemShut {NoStop}%
\bibitem [{\citenamefont {Waleffe}(1997)}]{waleffe1997}%
  \BibitemOpen
  \bibfield  {author} {\bibinfo {author} {\bibfnamefont {F.}~\bibnamefont
  {Waleffe}},\ }\bibfield  {title} {\bibinfo {title} {On a self-sustaining
  process in shear flows},\ }\href@noop {} {\bibfield  {journal} {\bibinfo
  {journal} {Phys. Fluids}\ }\textbf {\bibinfo {volume} {9}},\ \bibinfo {pages}
  {883} (\bibinfo {year} {1997})}\BibitemShut {NoStop}%
\bibitem [{\citenamefont {Hamba}(2018)}]{hamba2018}%
  \BibitemOpen
  \bibfield  {author} {\bibinfo {author} {\bibfnamefont {F.}~\bibnamefont
  {Hamba}},\ }\bibfield  {title} {\bibinfo {title} {Turbulent energy density in
  scale space for inhomogeneous turbulence},\ }\href@noop {} {\bibfield
  {journal} {\bibinfo  {journal} {J. Fluid Mech.}\ }\textbf {\bibinfo {volume}
  {842}},\ \bibinfo {pages} {532} (\bibinfo {year} {2018})}\BibitemShut
  {NoStop}%
\bibitem [{\citenamefont {Leith}(1990)}]{leith1990}%
  \BibitemOpen
  \bibfield  {author} {\bibinfo {author} {\bibfnamefont {C.~E.}\ \bibnamefont
  {Leith}},\ }\bibfield  {title} {\bibinfo {title} {Stochastic backscatter in a
  subgrid-scale model: {Plane} shear mixing layer},\ }\href@noop {} {\bibfield
  {journal} {\bibinfo  {journal} {Phys. Fluids A}\ }\textbf {\bibinfo {volume}
  {2}},\ \bibinfo {pages} {297} (\bibinfo {year} {1990})}\BibitemShut {NoStop}%
\bibitem [{\citenamefont {Langford}\ and\ \citenamefont
  {Moser}(1999)}]{lm1999}%
  \BibitemOpen
  \bibfield  {author} {\bibinfo {author} {\bibfnamefont {J.~A.}\ \bibnamefont
  {Langford}}\ and\ \bibinfo {author} {\bibfnamefont {R.~D.}\ \bibnamefont
  {Moser}},\ }\bibfield  {title} {\bibinfo {title} {Optimal {LES} formulations
  for isotropic turbulence},\ }\href@noop {} {\bibfield  {journal} {\bibinfo
  {journal} {J. Fluid Mech.}\ }\textbf {\bibinfo {volume} {398}},\ \bibinfo
  {pages} {321} (\bibinfo {year} {1999})}\BibitemShut {NoStop}%
\bibitem [{\citenamefont {Clark}\ \emph {et~al.}(1979)\citenamefont {Clark},
  \citenamefont {Ferziger},\ and\ \citenamefont {Reynolds}}]{clarketal1979}%
  \BibitemOpen
  \bibfield  {author} {\bibinfo {author} {\bibfnamefont {R.~A.}\ \bibnamefont
  {Clark}}, \bibinfo {author} {\bibfnamefont {J.~H.}\ \bibnamefont
  {Ferziger}},\ and\ \bibinfo {author} {\bibfnamefont {W.~C.}\ \bibnamefont
  {Reynolds}},\ }\bibfield  {title} {\bibinfo {title} {Evaluation of
  subgrid-scale models using an accurately simulated turbulent flow},\
  }\href@noop {} {\bibfield  {journal} {\bibinfo  {journal} {J. Fluid Mech.}\
  }\textbf {\bibinfo {volume} {91}},\ \bibinfo {pages} {1} (\bibinfo {year}
  {1979})}\BibitemShut {NoStop}%
\bibitem [{\citenamefont {Germano}\ \emph {et~al.}(1991)\citenamefont
  {Germano}, \citenamefont {Piomelli}, \citenamefont {Moin},\ and\
  \citenamefont {Cabot}}]{germanoetal1991}%
  \BibitemOpen
  \bibfield  {author} {\bibinfo {author} {\bibfnamefont {M.}~\bibnamefont
  {Germano}}, \bibinfo {author} {\bibfnamefont {U.}~\bibnamefont {Piomelli}},
  \bibinfo {author} {\bibfnamefont {P.}~\bibnamefont {Moin}},\ and\ \bibinfo
  {author} {\bibfnamefont {W.~H.}\ \bibnamefont {Cabot}},\ }\bibfield  {title}
  {\bibinfo {title} {A dynamic subgrid-scale eddy viscosity model},\
  }\href@noop {} {\bibfield  {journal} {\bibinfo  {journal} {Phys. Fluids A}\
  }\textbf {\bibinfo {volume} {3}},\ \bibinfo {pages} {1760} (\bibinfo {year}
  {1991})}\BibitemShut {NoStop}%
\bibitem [{\citenamefont {Lilly}(1992)}]{lilly1992}%
  \BibitemOpen
  \bibfield  {author} {\bibinfo {author} {\bibfnamefont {D.~K.}\ \bibnamefont
  {Lilly}},\ }\bibfield  {title} {\bibinfo {title} {A proposed modification of
  the {Germano} subgrid-scale closure method},\ }\href@noop {} {\bibfield
  {journal} {\bibinfo  {journal} {Phys. Fluids A}\ }\textbf {\bibinfo {volume}
  {4}},\ \bibinfo {pages} {633} (\bibinfo {year} {1992})}\BibitemShut {NoStop}%
\bibitem [{\citenamefont {Moser}\ \emph {et~al.}(1999)\citenamefont {Moser},
  \citenamefont {Kim},\ and\ \citenamefont {Mansour}}]{mkm1999}%
  \BibitemOpen
  \bibfield  {author} {\bibinfo {author} {\bibfnamefont {R.~D.}\ \bibnamefont
  {Moser}}, \bibinfo {author} {\bibfnamefont {J.}~\bibnamefont {Kim}},\ and\
  \bibinfo {author} {\bibfnamefont {N.~N.}\ \bibnamefont {Mansour}},\
  }\bibfield  {title} {\bibinfo {title} {Direct numerical simulation of
  turbulent channel up to ${Re}_\tau = 590$},\ }\href@noop {} {\bibfield
  {journal} {\bibinfo  {journal} {Phys. Fluids}\ }\textbf {\bibinfo {volume}
  {11}},\ \bibinfo {pages} {943} (\bibinfo {year} {1999})}\BibitemShut
  {NoStop}%
\bibitem [{\citenamefont {Abe}\ \emph {et~al.}(2009)\citenamefont {Abe},
  \citenamefont {Antonia},\ and\ \citenamefont {Kawamura}}]{abeetal2009}%
  \BibitemOpen
  \bibfield  {author} {\bibinfo {author} {\bibfnamefont {H.}~\bibnamefont
  {Abe}}, \bibinfo {author} {\bibfnamefont {R.~A.}\ \bibnamefont {Antonia}},\
  and\ \bibinfo {author} {\bibfnamefont {H.}~\bibnamefont {Kawamura}},\
  }\bibfield  {title} {\bibinfo {title} {Correlation between small-scale
  velocity and scalar fluctuations in a turbulent channel flow},\ }\href@noop
  {} {\bibfield  {journal} {\bibinfo  {journal} {J. Fluid Mech.}\ }\textbf
  {\bibinfo {volume} {627}},\ \bibinfo {pages} {1} (\bibinfo {year}
  {2009})}\BibitemShut {NoStop}%
\end{thebibliography}%

\end{document}